\documentclass[12pt]{article}
\usepackage{geometry}
\usepackage{a4}
\usepackage{graphicx}
\usepackage{epsf}
\usepackage{amsmath}
\usepackage{amssymb}
\usepackage{cite}
\newcommand{\be}{\begin{equation}}
\newcommand{\ee}{\end{equation}}

\newcommand{\Rmnum}[1]{\expandafter\@slowromancap\romannumeral #1@}
\newcommand{\bea}{\begin{eqnarray}}
\newcommand{\eea}{\end{eqnarray}}

\begin{document}
\def\A{{\mathbb{A}}}
\def\B{{\mathbb{B}}}
\def\C{{\mathbb{C}}}
\def\R{{\mathbb{R}}}
\def\s{{\mathbb{S}}}
\def\T{{\mathbb{T}}}
\def\Z{{\mathbb{Z}}}
\def\W{{\mathbb{W}}}
\begin{titlepage}
\title{Generalized Holographic Superconductors with Higher Derivative Couplings}
\author{}
\date{
Anshuman Dey, Subhash Mahapatra, Tapobrata Sarkar
\thanks{\noindent E-mail:~ deyanshu, subhmaha, tapo @iitk.ac.in}
\vskip0.4cm
{\sl Department of Physics, \\
Indian Institute of Technology,\\
Kanpur 208016, \\
India}}
\maketitle
\abstract{
\noindent
We introduce and study generalized holographic superconductors with higher derivative couplings between the field strength tensor and a complex scalar field, in
four dimensional AdS black hole backgrounds. We study this theory in the probe limit, as well as with backreaction. There are multiple tuning parameters in the theory, and with
two non-zero parameters, we show that the theory has a rich phase structure, and in particular,  the transition from the normal to the superconducting phase can be tuned to be
of first order or of second order within a window of one of these. This is established numerically as well as by computing the free energy of the boundary theory. 
We further present analytical results for the critical temperature of the model, and compare these with numerical analysis. 
Optical properties of this system are also studied numerically in the probe limit, and our results show evidence for negative refraction at low frequencies. 
}
\end{titlepage}

\section{Introduction}
One of the most interesting developments in String theory in the last two decades is the Maldacena conjecture, broadly known as the AdS/CFT correspondence \cite{Maldacena},\cite{Klebanov},\cite{Witten}.
This relates a classical theory of gravity in $d+1$ dimensional AdS space-time to a conformal field theory (CFT) in one lower dimension, a theory which lives at the boundary of the AdS space.
This is also known as the strong-weak coupling duality in the sense that the strong coupling behavior of the boundary CFT can be described by classical gravity in AdS space.
This duality offers a new theoretical tool using which one can gain insight into the physics of strongly coupled systems via classical gravity in AdS space, which otherwise may not be tractable.

Indeed, in the last few years, AdS/CFT has been widely used to study strongly coupled material systems that arise in condensed matter theories, and several new and interesting insights have been obtained from
such analyses. In particular, a holographic description of the phenomenon of superconductivity has been established by the works of \cite{Gubser},\cite{Hartnoll} (for general reviews of the subject,
see \cite{Horowitz}-\cite{Hartnollrev}). The essential idea here  is to consider an Abelian Higgs model in the background of an AdS black hole and spontaneously break the gauge symmetry.
This is analogous to the Higgs mechanism in AdS space. The minimum ingredients required for this process are a charged AdS black hole, a complex scalar field and a Maxwell field with minimal
interaction between them. It was shown by Gubser \cite{Gubser} that RN-AdS black holes become unstable at low temperatures, developing scalar hair. In the dual CFT side, this hairy AdS black hole
is indicative of the condensation of a charged scalar. In the models of \cite{Gubser},\cite{Hartnoll}, a non zero value of the charged condensate implies a spontaneous breaking of a global $U(1)$ symmetry 
in the absence of sources, and indicates a second order phase transition from a normal to a superconducting phase. Such symmetry breaking can also result in first order phase transitions, as we will
see in sequel. 

Soon after the realization of superconductors from holography, the authors of \cite{Franco} generalized this model by considering a non minimal interaction between the complex scalar and the Maxwell field.
They showed the normal-superconductor phase transition by spontaneously breaking the global $U(1)$ symmetry via a St\"{u}ckelberg mechanism. Interestingly, their work points to the existence of
a first order phase transition, and a metastable region in the superconducting phase. This is phenomenologically important, since there are a large number of superconductors which 
show first order phase transitions (see, e.g. \cite{super1},\cite{super2}), although predictions on holographic superconductors via AdS/CFT are still far from being tested in the laboratory.

In \cite{Siopsis}, the authors introduced a higher-derivative coupling between the field strength tensor and the scalar field via a coupling constant $\eta$. 
They studied the effect of $\eta$ and an external magnetic field on droplet formation in holographic superconductors in the probe limit. The motivation for considering higher derivative terms 
in the physics of holographic superconductors is that these appear in a top-down approach in truncations of String Theory or M-theory. The approach of \cite{Siopsis} is 
however phenomenological in nature, in the sense that a higher derivative correction is proposed, rather than derived from a full String Theory. The importance of such an approach 
lies in the fact that it often provides useful insights into the nature of strongly coupled systems that might prove to be beneficial in realistic model building. For a top-down approach 
to similar issues, see \cite{bobev1}, \cite{bobev2}. 

It is important and interesting to extend existing results to other phenomenologically interesting situations. For example, a concrete question one might ask is the nature of phase
transitions in generalized holographic superconductors in the presence of higher derivative couplings. This is expected to provide a richer phase structure compared to a minimally coupled
holographic superconductor, since we have  multiple tunable parameters, and should be interesting to analyze. A further issue that one might investigate is the behavior of the
response functions of this strongly coupled boundary theory. In particular, one can analyze the refractive index of the theory, and determine whether the generalized holographic superconductor
with higher derivative couplings show negative refraction. \footnote{We proceed with the usual caveat in mind, namely that there are no dynamical photons at the boundary, and we assume
that our theory is weakly coupled to a dynamical photon \cite{policastro}.}

\begin{figure}[t!]
\centering
\includegraphics[width=2.7in,height=2.3in]{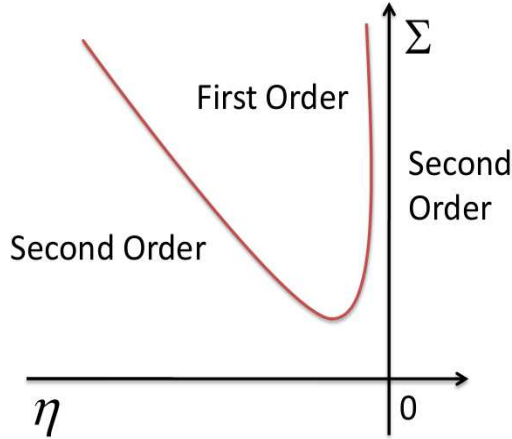}
\caption{Order of the phase transitions as a function of the model parameters.}
\label{fullphasestructure}
\end{figure}

In this paper, we undertake these analyses, and the purpose of this work is to understand 
a generalized holographic superconductor with a higher derivative coupling, both in the probe limit and with backreaction. Our main results are summarized as follows : \\
$\bullet$ We find that in the parameter space of our theory (denoted by $\eta$ and $\Sigma$ in sequel), the transition from normal to superconducting phase can be 
of first order within a window of parameter values, and is otherwise of second order. This is schematically depicted in fig.(\ref{fullphasestructure}).\\
$\bullet$ It is shown that the ratio of the energy gap to the critical temperature can change considerably by inclusion of higher derivative terms and is further enhanced by including back reaction.\\
$\bullet$ By studying optical properties of the system, we find that the refractive index can be negative in the probe limit (a similar result was found in \cite{mps}). We also find indication
that the imaginary part of the magnetic permeability might be positive even in the probe limit due to higher derivative corrections. However more analysis is required to establish
these results.

This paper is organized as follows. In section 2, we will introduce the model with higher derivative couplings and set up the basic notations and conventions to be followed in the
rest of the paper. We will generalize this model by introducing 
non minimal couplings between the field strength tensor and the scalar field. We will then show that our boundary system exhibits normal-superconductor type phase transitions at a critical 
temperature $T_c$ and that the nature of the phase transition depends on the coupling parameter $\eta$. In section 3, by introducing gauge field perturbations, we will study 
the effect of $\eta$ and other parameters in the model on the optical conductivity. In section 4, we will calculate $T_c$ analytically both in probe limit and with backreaction. In section 5,
we study optical properties of our model, and show that with higher derivative terms, the refractive index can in general be negative even in the probe limit. Finally, section 6 concludes 
this paper with some discussions and directions for future research.

\section{\textbf{Generalized Holographic Superconductors with a Higher Derivative Coupling}}

In this section we will introduce the basic setup to construct holographic superconductors with higher derivative couplings. As pointed out in \cite{Gubser} \cite{Hartnoll},
the minimum constituents required to make a boundary theory superconducting are a $U(1)$ gauge field and a complex scalar field in an AdS black hole background.
By adding higher derivative couplings between the field strength tensor and the scalar field on this set up \cite{Siopsis}, we can start with the following action
\begin{eqnarray}
\textit{S}= \int \mathrm{d^{4}}x\! \sqrt{-g}\biggl[\frac{1}{2\kappa^{2}}\biggl(R+\frac{6}{L^{2}}\biggr)
-\frac{1}{4}\textit{F}_{\mu\nu}\textit{F}^{\mu\nu}-\frac{1}{2}|\textit{D}_{\mu}\tilde{\Psi}|^{2} &\nonumber \\ -\frac{1}{2}m^{2}|\tilde{\Psi}|^{2}
-\frac{\eta}{2}|\textit{F}_{\mu\nu}\textit{D}^{\nu}\tilde{\Psi}|^{2}\biggl] \,
\label{action}
\end{eqnarray}
where $\kappa$ is related to the Newton's constant in four dimensions, $F=dA$ and $D_{\mu}=\partial_{\mu}-i q A_{\mu}$. $\tilde{\Psi}$ is the complex scalar field with mass $m$
and charge $q$, and $L$ is the AdS length scale. The last term in eq.(\ref{action}) describes the higher derivative interaction between the field strength tensor and the scalar field.  Re-writing the
charged scalar field $\tilde{\Psi}$ as $\tilde{\Psi}= \Psi  e^{i \alpha}$, the above action can be cast as
\begin{eqnarray}
\textit{S}= \int \mathrm{d^{4}}x\! \sqrt{-g}\biggl[\frac{1}{2\kappa^{2}}\biggl(R+\frac{6}{L^{2}}\biggr)-\frac{1}{4}\textit{F}_{\mu\nu}\textit{F}^{\mu\nu}-\frac{(\partial_{\mu}\Psi)^2}{2}-
\frac{m^2\Psi^2}{2} &\nonumber \\ -\frac{\eta}{2}\textit{F}_{\mu\nu}\partial^{\nu}\Psi \textit{F}^{\mu\sigma}\partial_{\sigma}\Psi -\frac{\Psi^2(\partial\alpha-q A)^2}{2}- 
\frac{\eta}{2}\Psi^2\biggl(\textit{F}^{\mu\nu}(\partial_{\nu}\alpha-q A_{\nu})\biggr)^{2} \biggr] \,
\label{actionpsi2}
\end{eqnarray}
where the $U(1)$ gauge symmetry is now given by $A_{\mu}\rightarrow A_{\mu}+\partial_{\mu}\lambda$, along with $\alpha\rightarrow \alpha+q \lambda$. In this notation, both $\alpha$ and $\Psi$ are real.
Now, one can generalize the above action in a gauge invariant way by replacing $|\Psi|^2$ by analytic functions of $\Psi$. In particular, we can do this by introducing two different
functions ${\rm G}(\Psi)$ and ${\rm K}(\Psi)$ in the second line of eq.(\ref{actionpsi2}), and generalize the action of eq.(\ref{actionpsi2}) as 
\begin{eqnarray}
\textit{S}= \int \mathrm{d^{4}}x\! \sqrt{-g}\biggl[\frac{1}{2\kappa^{2}}\biggl(R+\frac{6}{L^{2}}\biggr)-\frac{1}{4}\textit{F}_{\mu\nu}\textit{F}^{\mu\nu}-\frac{(\partial_{\mu}\Psi)^2}{2} 
-\frac{\eta}{2}\textit{F}_{\mu\nu}\partial^{\nu}\Psi \textit{F}^{\mu\sigma}\partial_{\sigma}\Psi &\nonumber \\ -\frac{m^2\Psi^2}{2} -\frac{|\textrm{G}(\Psi)|(\partial\alpha-q A)^2}{2}
- \frac{\eta}{2}|\textrm{K}(\Psi)|\biggl(\textit{F}^{\mu\nu}(\partial_{\nu}\alpha-q A_{\nu})\biggr)^{2} \biggr]
\label{action3}
\end{eqnarray}
For $\eta=0$, this model reduces to the one first studied in \cite{Franco} in the probe limit, which was then subsequently generalized to include back reaction in \cite{Yuan}, \cite{QPan}.
It is known that by considering different forms of $\textrm{G}(\Psi)$, one can tune the order of the phase transition as well as change the mean field exponents. Here we employ
a different method of generalizing the model by introducing a second functional form of $\Psi$ with the higher derivative correction. This is the novelty of our model, and as shown
in sequel, gives rise to several interesting results. 

A word about the probe limit and backreaction  in our model is in order. If one scales $\Psi\rightarrow \Psi/q$ and $A_{\mu}\rightarrow A_{\mu}/q$, then the action of eq.(\ref{action3}) 
can  be written as 
 \begin{eqnarray}
\textit{S}= \int \mathrm{d^{4}}x\! \sqrt{-g}\biggl[\frac{1}{2\kappa^{2}}\biggl(R+\frac{6}{L^2}\biggr)-\frac{1}{4 q^2}\textit{F}_{\mu\nu}\textit{F}^{\mu\nu}
-\frac{(\partial_{\mu}\Psi)^2}{2 q^2} -\frac{\eta}{2 q^4}\textit{F}_{\mu\nu}\partial^{\nu}\Psi \textit{F}^{\mu\sigma}\partial_{\sigma}\Psi &\nonumber \\ 
-\frac{m^2\Psi^2}{2 q^2} -\frac{1}{2 q^2}|{\rm G}(\Psi)|(\partial\alpha-A)^2  - \frac{\eta}{2 q^4}|{\rm K}(\Psi)|\biggl(\textit{F}^{\mu\nu}(\partial_{\nu}\alpha-  A_{\nu})\biggr)^{2} \biggr]
\end{eqnarray} 
where the scaling pattern of ${\rm G}(\Psi)$ and ${\rm K}(\Psi)$ is $q^{-2}$ since $\Psi^2$ is the leading terms in these expressions.\footnote{In all models of generalized holographic
superconductors, there is a mass scale arising out of dimensional analysis that needs to be set to unity \cite{Franco}. This will be clear from eq.(\ref{generalizedterms}).}
From the matter part of the action, we see that each $\eta$ term comes with a factor of $1/q^4$ while terms without $\eta$ come with a factor of $1/q^2$.
The probe limit is normally defined as $\kappa^2/q^2\rightarrow0$. In this paper, we will consider $\kappa^2 \to 0$ with $q^2=1$ as the probe limit. In the other alternative, where we can
set $\kappa^2 = 1$ and take $q^2 \to \infty$, the terms involving $\eta$ will drop out, and hence this is not very useful for us. Importantly, for most cases, 
the probe limits in the two approaches are equivalent. However, in models with higher derivative couplings, they are not.
Also, there are two ways to go away from the probe limit, i.e consider back reaction. First, by taking $\kappa^2 = 1$ and simultaneously  choosing a finite value of $q^2$, an approach used in \cite{Hartnollcorev}.
Alternatively  one can consider backreaction by setting $q^2=1$ and work with the finite variable $\kappa^2$ (see, e.g \cite{amariti}). In this paper we use the latter approach. 

Now by varying the action, we find the equations of motion (EOM) for the scalar and the gauge field as,
\begin{eqnarray}
&& \frac{1}{\sqrt{-g}}\partial_{\mu}\biggl[\sqrt{-g}(\partial^{\mu}\Psi-\eta\textit{F}^{\mu\nu}\textit{F}_{\nu\sigma}\partial^{\sigma}\Psi)\biggr] 
-\frac{\eta}{2}\biggl(\textit{F}_{\rho\lambda}(\partial^{\lambda}\alpha-A^{\lambda})\biggr)^{2}\frac{d\textrm{K}(\Psi)}{d\Psi} \nonumber \\ && - m^{2}\Psi 
- \frac{(\partial\alpha-A)^{2}}{2} \frac{d\textrm{G}(\Psi)}{d\Psi}=0
\end{eqnarray}
\begin{eqnarray}
&& \partial_{\mu}(\sqrt{-g}\textit{F}^{\mu\nu}) + \sqrt{-g}\biggl(\textrm{G}(\Psi)(\partial^{\nu}\alpha-A^{\nu})-\eta\textrm{K}(\Psi)\textit{F}^{\nu\rho}\textit{F}_{\rho\sigma} 
(\partial^{\sigma}\alpha-A^{\sigma})\biggr) \nonumber \\ && + \partial_{\mu}\biggl(\eta \sqrt{-g}(\textit{F}^{\mu\sigma}\partial_{\sigma}\Psi\partial^{\nu}\Psi -
\textit{F}^{\nu\sigma}\partial_{\sigma}\Psi\partial^{\mu}\Psi)\biggr) +\partial_{\mu}\biggl(\sqrt{-g}\eta\textrm{K}(\Psi) \nonumber \\ && \biggl(\textit{F}^{\mu\sigma}
(\partial_{\sigma}\alpha-A_{\sigma})(\partial^{\nu}\alpha-A^{\nu})-\textit{F}^{\nu\sigma}(\partial_{\sigma}\alpha-A_{\sigma})(\partial^{\mu}\alpha-A^{\mu})\biggr)\biggr)=0
\end{eqnarray}
\begin{eqnarray}
\partial_{\mu}\biggl[\sqrt{-g}\biggl(\textrm{G}(\Psi)(\partial^{\mu}\alpha-A^{\mu})-\eta\textrm{K}(\Psi)\textit{F}^{\mu\rho}\textit{F}_{\rho\sigma} (\partial^{\sigma}\alpha-A^{\sigma})\biggr)\biggr]=0
\end{eqnarray}
where we have set $q=1$. Also, the Einstein's EOM reads 
\begin{eqnarray}
&& \frac{1}{2\kappa^{2}}\biggl(R_{\mu\nu}-\frac{1}{2}g_{\mu\nu}R-\frac{3g_{\mu\nu}}{L^{2}}\biggr) + \frac{1}{8} g_{\mu\nu} \textit{F}^{\ 2}-
\frac{1}{2}\textit{F}_{\mu\lambda}\textit{F}_{\nu}^{ \ \ \lambda}+\frac{1}{4}g_{\mu\nu}m^{2}\Psi^{2}+
\frac{1}{4}g_{\mu\nu}(\partial\Psi)^{2} \nonumber \\ && -\frac{1}{2}\partial_{\mu}\Psi\partial_{\nu}\Psi
+\frac{1}{4}g_{\mu\nu}\textrm{G}(\Psi)(\partial\alpha-A)^{2} - \frac{1}{2}\textrm{G}(\Psi)(\partial_{\nu}\alpha-A_{\nu})(\partial_{\mu}\alpha-A_{\mu}) \nonumber \\ &&
- \frac{\eta}{2}\biggl(-\frac{1}{2}g_{\mu\nu}(\textit{F}_{\rho\lambda}\partial^{\lambda}\Psi)^{2}
+\textit{F}_{\mu\lambda}\partial^{\lambda}\Psi\textit{F}_{\nu\rho}\partial^{\rho}\Psi - \partial_{\mu}\Psi\textit{F}_{\nu}^ {\ \ \rho}\textit{F}_{\rho\lambda}\partial^{\lambda}\Psi - 
\partial_{\nu}\Psi\textit{F}_{\mu}^{\ \ \rho}\textit{F}_{\rho\lambda}\partial^{\lambda}\Psi \biggr) \nonumber \\ &&
-\frac{\eta}{2}\textrm{K}(\Psi)\biggl(-\frac{1}{2}g_{\mu\nu}\biggl(\textit{F}_{\rho\lambda}(\partial^{\lambda}\alpha-A^{\lambda})\biggr)^{2}  -\textit{F}_{\mu\rho}(\partial_{\nu}\alpha-
A_{\nu})\textit{F}^{\rho\lambda}(\partial_{\lambda}\alpha-A_{\lambda}) \nonumber \\ &&
+\textit{F}_{\mu\rho}(\partial^{\rho}\alpha-A^{\rho})\textit{F}_{\nu\lambda}(\partial^{\lambda}\alpha-A^{\lambda})
-(\partial_{\mu}\alpha-A_{\mu})\textit{F}_{\nu}^{ \ \ \rho}\textit{F}_{\rho\lambda}(\partial^{\lambda}\alpha-A^{\lambda})\biggr)   =0
\label{einsteineom}
\end{eqnarray}
In the rest of the paper, we will use the gauge $\alpha=0$. Since we are interested in (planar) charged, hairy black hole like solutions including the backreactions of the
gauge and scalar fields, we consider the following ansatz :
\begin{eqnarray}
\textit{d}s^{2}=-g(r)e^{-\chi(r)}dt^{2}+r^{2}(dx^{2}+dy^{2})+\frac{dr^{2}}{g(r)}
\label{metric}
\end{eqnarray}
\begin{equation}
\Psi=\Psi(r),~~~A=\Phi(r)dt
\end{equation}
The Hawking temperature of the black hole is given by
\begin{equation}
T_{H}=\frac{g'(r)e^{-\chi(r)/2}}{4\pi}|_{r=r_{h}}
\label{Hawking}
\end{equation}
where $r_{h}$ is the radius of event horizon which is given by the solution of $g(r_{h})=0$. In the rest of the paper, $g(r)$ should be understood as the coefficient of
$dt^{2}$ in eq. (\ref{metric}) and should not be confused with the determinant of the metric. With the ansatz of eq.(\ref{metric}), one can show that the scalar and the gauge field EOMs reduce to
\begin{eqnarray}
\Psi '' \left(1-\eta  e^{\chi } \Phi'^2\right)+ \frac{ \Phi ^2 e^{\chi }}{2g^2}\frac{d\textrm{G}(\Psi)}{d\Psi}  -\frac{\eta   \Phi ^2 e^{2 \chi } \Phi '^2}{2 g^2}
\frac{d\textrm{K}(\Psi)}{d\Psi}+ \Psi '\left( \frac{2}{r}+\frac{g'}{g}-\frac{\chi '}{2}\right) &\nonumber \\  -\frac{m^2 \Psi }{g} + \Psi ' \left(-\frac{\eta  e^{\chi } g' \Phi '^2}{g}-\frac{1}{2}
\eta  e^{\chi } \Phi '^2 \chi '-\frac{2 \eta  e^{\chi} \Phi '^2}{r}-2 \eta  e^{\chi } \Phi ' \Phi ''\right)=0
\label{scalareom}
\end{eqnarray}
\begin{eqnarray}
&& \Phi '' \left(-\frac{\eta  \textrm{K}(\Psi) \Phi^2 e^{\chi }}{g}+\eta  g \Psi '^2+1\right) -\Phi
\left(\frac{\textrm{G}(\Psi)}{g}+\frac{\eta  \textrm{K}(\Psi) e^{\chi } \Phi'^2}{g}\right) \nonumber \\ &&  +\Phi ' \left(\eta  g' \Psi'^2+\frac{1}{2} \eta
g \chi ' \Psi '^2+\frac{2 \eta  g \Psi '^2}{r}+2 \eta  g \Psi ' \Psi ''+\frac{\chi '}{2}+\frac{2}{r}\right) \nonumber \\ && + \eta \Phi ^2 \Phi ' \left(\frac{  \textrm{K}(\Psi) e^{\chi } g'}{g^2}-
\frac{e^{\chi } \textrm{K}(\Psi)'}{g}-\frac{3 \textrm{K}(\Psi) e^{\chi } \chi '}{2 g}-\frac{2 \textrm{K}(\Psi) e^{\chi }}{r g}\right) =0
\label{gaugeeom}
\end{eqnarray}
Similarly, the $tt$ and the $rr$ components of Einstein equations give
\begin{eqnarray}
&& g'+2\kappa^{2} r \biggl(\frac{ \textrm{G}(\Psi) \Phi ^2 e^{\chi }}{4 g}-\frac{3   \eta  \textrm{K}(\Psi) \Phi ^2 e^{2 \chi } \Phi '^2}{4
   g}+\frac{1}{4} \eta  g e^{\chi } \Phi '^2 \Psi '^2 + \frac{1}{4} g \Psi '^2 \nonumber \\ && +\frac{1}{4} m^2 \Psi^2+\frac{1}{4} e^{\chi } \Phi '^2\biggr) -3 r +\frac{g}{r}=0
\label{tteinsteineom}
\end{eqnarray}
\begin{eqnarray}
2\kappa^{2}r\bigg(\frac{ \textrm{G}(\Psi) \Phi ^2 e^{\chi }}{2 g^2}-\frac{ \eta  \textrm{K}(\Psi) \Phi ^2 e^{2 \chi } \Phi '^2}{2 g^2}-\frac{1}{2}  \eta  e^{\chi } \Phi '^2 \Psi '^2+\frac{1}{2}  \Psi '^2\biggr)+ \chi '=0,
\label{rreinsteineom}
\end{eqnarray}
where a prime denotes a derivative with respect to $r$, and for simplicity we have explicitly suppressed the radial dependence of our variables.

We therefore have four coupled differential equations which need to be solved by appropriate boundary conditions. At the horizon, we impose the regularity conditions for $\Psi$ and $\Phi$,
where these fields behave as,
\begin{equation}
\Phi(r_{h})=0 ,\ \ \Psi'(r_{h})=\frac{m^2\Psi(r_{h})}{g'(r_{h})(1-\eta e^{\chi(r_{h})}\Phi'^{2}(r_{h}))}.
\label{horizon behavior}
\end{equation}
The first condition in the above equation is imposed by demanding a finite form for the gauge field, and the second condition comes from eq.(\ref{scalareom}).
Before discussing the asymptotic behavior of the fields at the boundary, we would like to mention here that there are three scale symmetries (which we will not explicitly write here) and one can use
these to set $r_{h}=1$, $L=1$ and $\chi =0$ at the boundary \cite{Hartnollcorev}. The asymptotic expressions of the fields near the boundary are given as
\begin{eqnarray}
\Phi=\mu-\frac{\rho}{r} +... , ~~ \Psi=\frac{\Psi_{-}}{r^{\lambda_{-}}}+\frac{\Psi_{+}}{r^{\lambda_{+}}} + ..., ~~ \chi\rightarrow 0, \ \ \ g\rightarrow r^{2}+...
\label{boundar behavior}
\end{eqnarray}
where $\lambda_{\pm}=\frac{3\pm\sqrt{9+4m^2}}{2}$. The third condition arises by the physical requirement that the boundary field theory temperature must be equal to the
Hawking temperature of the black hole. 

According to the AdS/CFT dictionary, in the mass range $-d^2/4<m^2<-d^2/4 + 1$, both $\Psi_{+}$ and $\Psi_{-}$ are normalizable at the boundary, and can act as the source or the vacuum
expectation value of the corresponding operators \cite{klebanovwitten},\cite{Robert}. In this paper we consider $\Psi_{-}$ as a source and put $\Psi_{-}=0$ as the boundary condition.  Also, in eq.(\ref{boundar behavior}),
$\mu$ and $\rho$ are the chemical potential and the charge density of the boundary theory respectively. From now on, we will consider a particular case where 
$m^{2}=-2$. Although $m^{2}$ is negative but it is above the Breitenlohner-Freedman (BF) bound $m^{2}_{BF}=-9/4$ \cite{Freedman}. For this value of $m^2$, 
we  have $\lambda_{\pm}=2,1$. Since we set $\Psi_{(1)} = 0$, the condensate of the scalar operator $O_{2}$ in the boundary theory dual to the scalar field is given by
\begin{equation}
<O_{2}>\sim{\Psi_{(2)}}
\label{O2}
\end{equation}
Due to the nonlinear nature of the coupled differential equations in (\ref{scalareom}) - (\ref{rreinsteineom}), it is difficult to find an analytic solution.
However, when the condensate is small, one can use series perturbations and matching method techniques to solve these equation analytically. Indeed, in section 4,
we will find the critical temperature $T_{c}$ analytically. However we would like to mention here that any analytic solution will be valid only near $T_{c}$
and away from it one has to resort to a full numerical analysis.

Now we present our numerical results. For numerical efficiency, it is convenient to use the coordinate $z=r_{h}/r$. With this choice, the boundary and
the horizon are located at $z=0$ and $z=1$ respectively. In  what follows, we will specialize to some particular forms of $\textrm{G}(\Psi)$ and $\textrm{K}(\Psi)$, namely
\begin{equation}
\textrm{G}(\Psi)=\Psi^{2}+\xi \Psi^{\theta},~~~\textrm{K}(\Psi)=\Psi^{2}+\Sigma \Psi^{\gamma}
\label{generalizedterms}
\end{equation}
There are a large number of parameters in this model. For simplicity, we will fix a few of them. In particular, we will set $$\gamma=4$$
We note here that one must choose $\gamma > 1, \theta >1$, since otherwise $\Psi$ appears in the denominator in some terms of the equations of motion, and hence a normal solution
$\Psi = 0$ is not allowed. For generalized superconductors, various values of $\gamma$ are allowed, $\gamma = 4$ is simply one particular choice. We have
explicitly checked for some other values of $\gamma$ that our results do not show any qualitative changes. 
Also, since we  are mostly interested in exploring the phase structure of the boundary system with respect to the higher derivative coupling parameter $\eta$ (and also with $\Sigma$),
we will set $\xi=0$, but again we have checked that a non zero value of $\xi$ do not change the results qualitatively. The complete phase structure of the model 
including $\xi$ will be studied elsewhere.

\begin{figure}[t!]
\begin{minipage}[b]{0.5\linewidth}
\centering
\includegraphics[width=2.7in,height=2.3in]{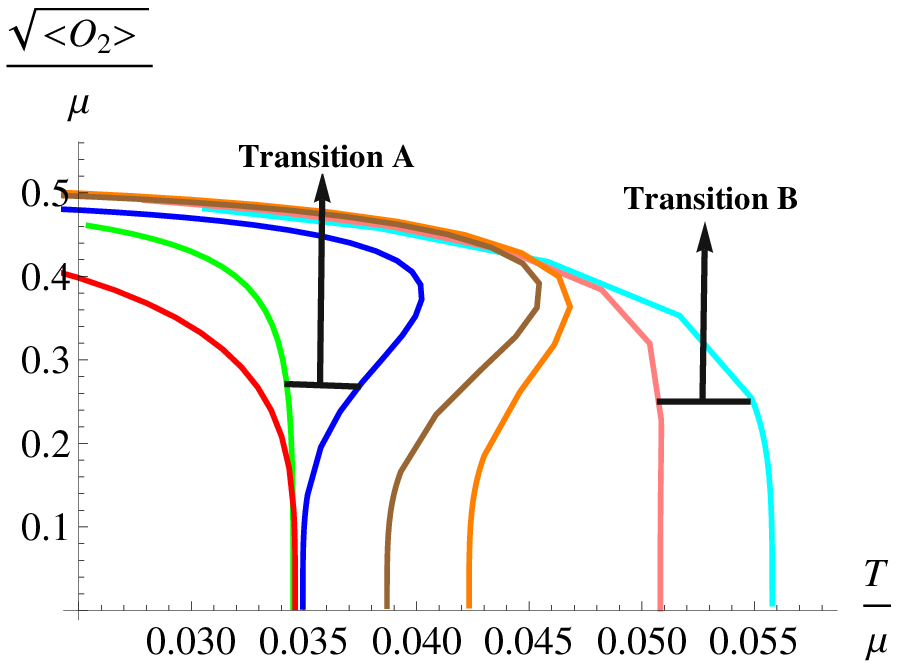}
\caption{ Condensate for fixed $\Sigma=5$ and $2\kappa^{2}=0.5$ for different values of $\eta$.}
\label{O2VsEtaAlpha0.5Sigma5}
\end{minipage}
\hspace{0.4cm}
\begin{minipage}[b]{0.5\linewidth}
\centering
\includegraphics[width=2.7in,height=2.3in]{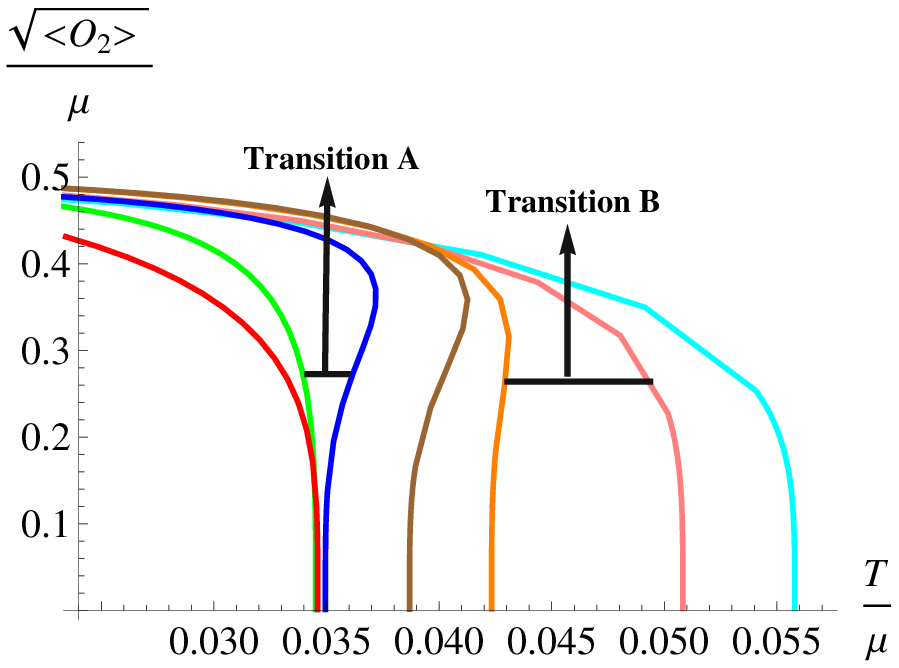}
\caption{Condensate for fixed $\Sigma=3$ and $2\kappa^{2}=0.5$ for different values of $\eta$.}
\label{O2VsEtaAlpha0.5Sigma3}
\end{minipage}
\end{figure}

\begin{figure}[t!]
\begin{minipage}[b]{0.5\linewidth}
\centering
\includegraphics[width=2.7in,height=2.3in]{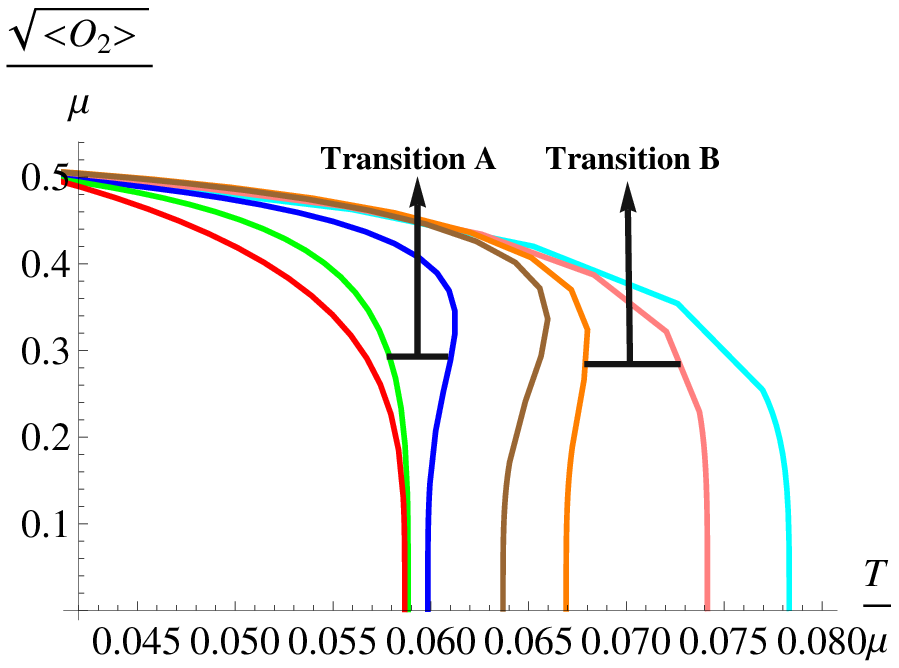}
\caption{ Condensate for fixed $\Sigma=5$ and $2\kappa^{2}=0$ for different values of $\eta$.}
\label{O2VsEtaAlpha0Sigma5}
\end{minipage}
\hspace{0.4cm}
\begin{minipage}[b]{0.5\linewidth}
\centering
\includegraphics[width=2.7in,height=2.3in]{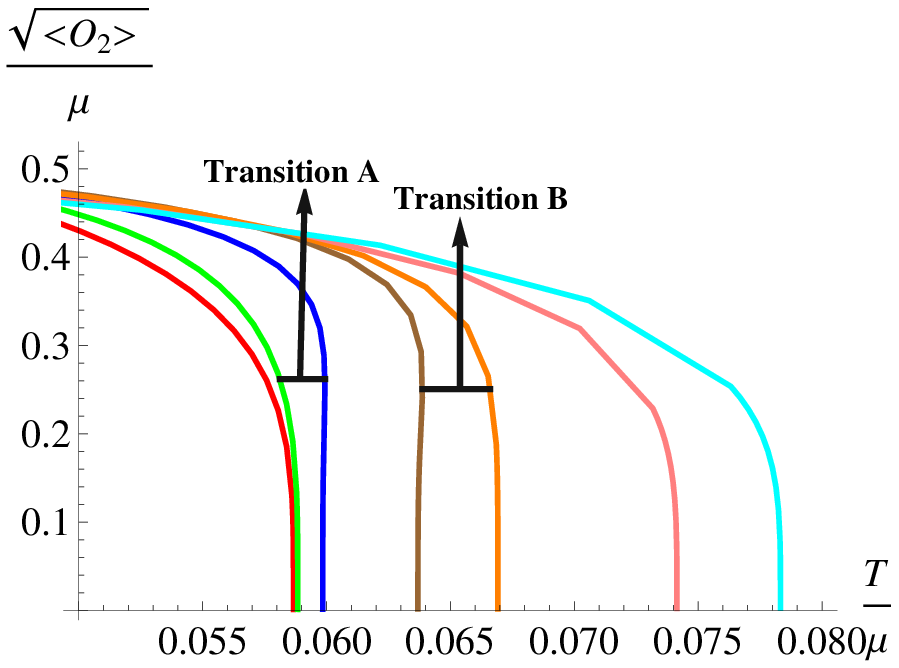}
\caption{Condensate for fixed $\Sigma=3$ and $2\kappa^{2}=0$ for different values of $\eta$.}
\label{O2VsEtaAlpha0Sigma3}
\end{minipage}
\end{figure}
\noindent

In figure (\ref{O2VsEtaAlpha0.5Sigma5}), we have shown the variation of $\sqrt{O_{2}}$, normalised by $\mu$, with respect to $T/\mu$ for different value of $\eta$.
Here, we have considered $2\kappa^{2}=0.5$ and $\Sigma=5$. The red, green, blue, brown, orange, pink and cyan curves correspond to $\eta$=0.01, -0.01, -0.1, -0.5, -1, -3, and -5
respectively\footnote{The same color coding is used in figure (\ref{O2VsEtaAlpha0.5Sigma3})-(\ref{O2VsEtaAlpha0Sigma3}), and we do not mention this in sequel.}.
We see from fig.(\ref{O2VsEtaAlpha0.5Sigma5}) that as $T/\mu$ falls below its critical value, the condensate develops a nonzero value, which indicates the onset of a superconducting
phase. Above this critical value, the system is in its normal phase. For smaller values of $\eta$, the transition point in  the $T/\mu$ axis, where the system goes into the superconducting phase,
increases (or $\mu_{c}$ decreases). This indicates that smaller values of $\eta$ make the condensate easier to form. 

Our main result here is in the behavior of the condensate for
different values of $\eta$. For $\eta$=0.01 and -0.01, the transition from the normal to the superconducting phase is of second order. But as we decrease $\eta$ from -0.01, first the transition
changes to first order and then again, for much smaller values $\eta$ to second order.
Indeed $\eta$=-0.1, -0.5, -1 and -3 produces first order phase transitions and a second order transition is seen for $\eta$=-5.
Phase transitions for $\eta<-5$ and $\eta>0.01$ are always of second order. This suggest the presence of two critical $\eta$s, say $\eta_{c1}$ and $\eta_{c2}$, at which the
nature of the phase transition changes.
\begin{figure}[h!]
\begin{minipage}[b]{0.5\linewidth}
\centering
\includegraphics[width=2.7in,height=2.3in]{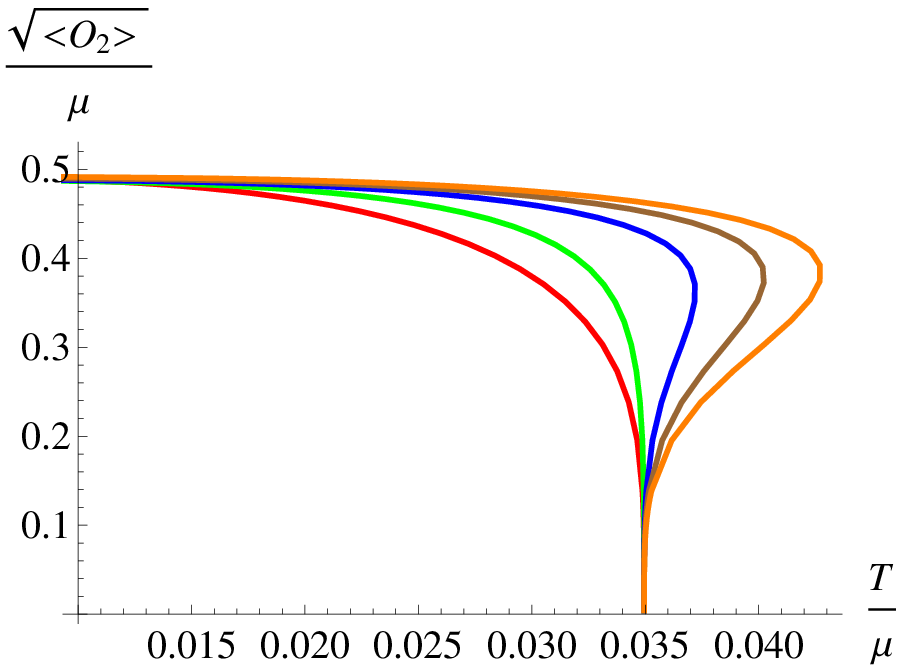}
\caption{ Condensate with $\eta=-0.1$, $2\kappa^{2}=0.5$ for different values of $\Sigma$.}
\label{O2VsSigmaAlpha0.5Eta-0.1}
\end{minipage}
\hspace{0.4cm}
\begin{minipage}[b]{0.5\linewidth}
\centering
\includegraphics[width=2.7in,height=2.3in]{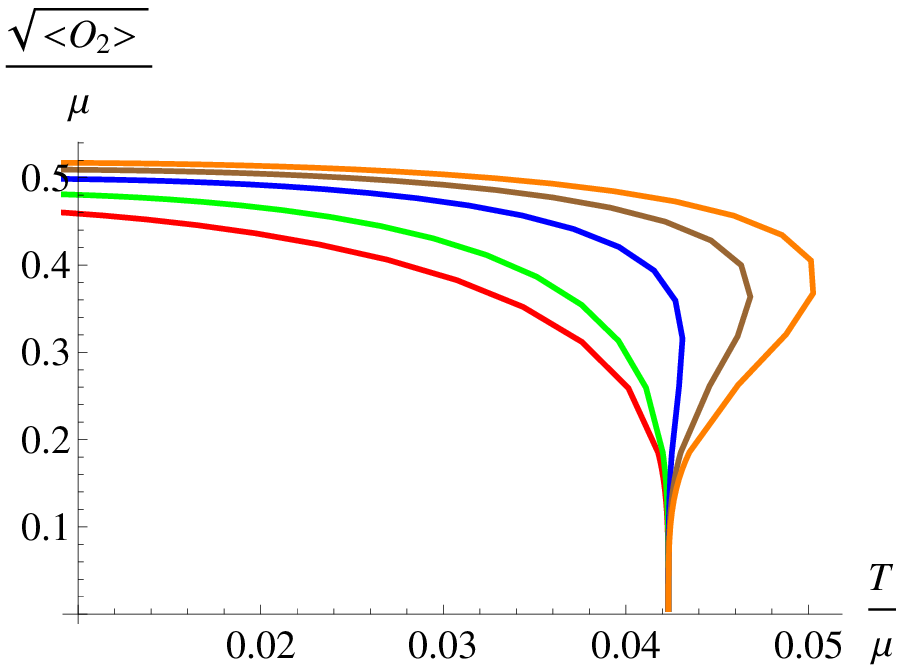}
\caption{Condensate with $\eta=-1$, $2\kappa^{2}=0.5$ for different values of $\Sigma$.}
\label{O2VsSigmaAlpha0.5Eta-1}
\end{minipage}
\end{figure}

\begin{figure}[t!]
\begin{minipage}[b]{0.5\linewidth}
\centering
\includegraphics[width=2.7in,height=2.3in]{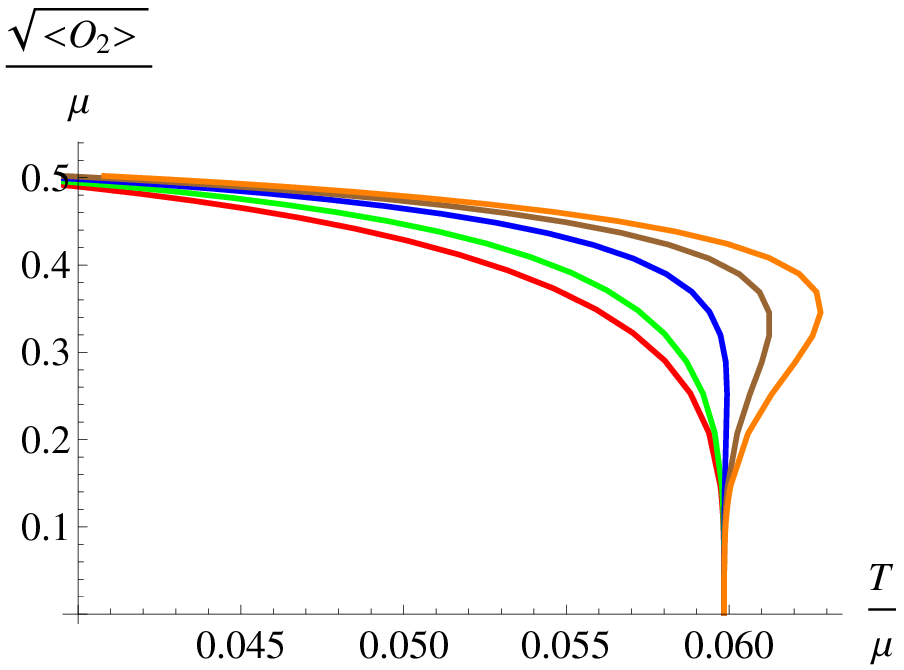}
\caption{ Condensate with $\eta=-0.1$, $2\kappa^{2}=0$ for different values of $\Sigma$.}
\label{O2VsSigmaAlpha0Eta-0.1}
\end{minipage}
\hspace{0.4cm}
\begin{minipage}[b]{0.5\linewidth}
\centering
\includegraphics[width=2.7in,height=2.3in]{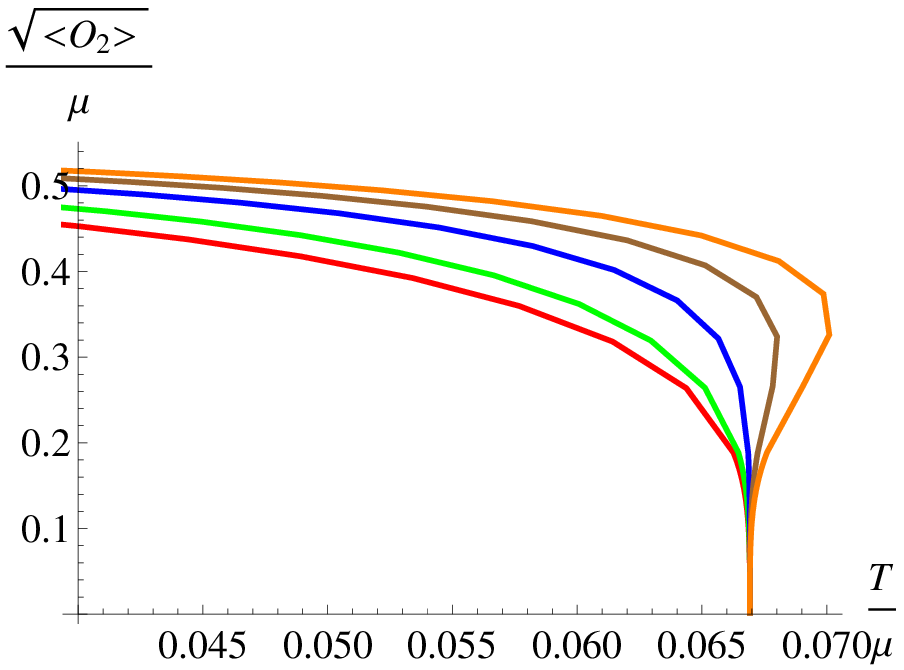}
\caption{Condensate with $\eta=-1$, $2\kappa^{2}=0$ for different values of $\Sigma$.}
\label{O2VsSigmaAlpha0Eta-1}
\end{minipage}
\end{figure}
\begin{eqnarray}
\left\{ \begin{array}{ll}
\eta> \eta_{c1}~~ ~~ {\rm second ~order} \\
\eta_{c2} < \eta < \eta_{c1}~~ ~~{\rm first ~ order}\\
\eta < \eta_{c2}~~ ~~ {\rm second ~ order}
\end{array} \right. \nonumber\\
\label{rangeEta}
\end{eqnarray}
We will call these as ``transition A'' and ``transition B'', respectively. Qualitatively, this was the behavior alluded to in the introduction, 
in figure(\ref{fullphasestructure}). We thus find that there exists a window in the parameter $\eta$ where the transition from the normal 
to the superconducting phase is of first order, and away from this window, the transition is of second order. To the best of our knowledge, such a window in the parameter
space has not appeared previously in the literature on holographic superconductors, and might be of significance in realistic systems. 

There is another observation regarding the magnitudes of condensate at low temperatures. As we decrease $\eta$, it first increases 
and then decreases. The decrease in the condensate value starts nearly after transition B \textit{i.e} $\eta < \eta_{c2}$. In figure (\ref{O2VsEtaAlpha0.5Sigma3}), we have chosen 
$2\kappa^{2}=0.5$ but $\Sigma=3$. Apart from a few differences, the results in figure (\ref{O2VsEtaAlpha0.5Sigma3}) are qualitatively similar to figure (\ref{O2VsEtaAlpha0.5Sigma5}). 
One important difference is the order of phase transition at $\eta=-3$ which is now second order as compared to first order in figure(\ref{O2VsEtaAlpha0.5Sigma5}). 
This indicates that $\eta_{c2}$ increases for smaller $\Sigma$ and therefore the window in $\eta$, where one gets first order phase transition, decreases by lowering $\Sigma$. 
In figure (\ref{O2VsEtaAlpha0.5Sigma5}) and (\ref{O2VsEtaAlpha0.5Sigma3}) we have used a fixed value of the
back reaction parameter $\kappa$. One can also vary $\kappa$ and obtain qualitatively similar results.

Next, in figs.(\ref{O2VsEtaAlpha0Sigma5}) and (\ref{O2VsEtaAlpha0Sigma3}), we have set $\kappa = 0$.
Comparing these with figs.(\ref{O2VsEtaAlpha0.5Sigma5}) and (\ref{O2VsEtaAlpha0.5Sigma3}), we find that $T_c$
decreases with $\kappa$. Consequently higher back reaction makes the condensate harder to form, which is consistent with existing results in the literature. Also, $\eta_{c2}$ decreases 
for smaller $\kappa$. Therefore, by analyzing figs.(\ref{O2VsEtaAlpha0.5Sigma5})-(\ref{O2VsEtaAlpha0Sigma3}), it is clear that the nature of the phase transition depends 
in a non trivial way on $\eta$ and $\Sigma$, along with parameter $\kappa$. Our numerical analysis also indicates that, $\Sigma$ and $\kappa$ play important roles in defining 
the range of the window in $\eta$ where one gets first order phase transitions.

Now we will analyze the theory for different values of $\Sigma$, for fixed $\eta$. This is shown in figs.(\ref{O2VsSigmaAlpha0.5Eta-0.1})-(\ref{O2VsSigmaAlpha0Eta-1}). 
Here, the red, green, blue, brown and orange curves correspondence to $\Sigma=0, 1, 3, 5 \ {\rm and} \ 7$, respectively. The nature of the normal-superconductor phase transition 
is different here. We find that there is a lower cut-off $\Sigma_{c}$ above which the transition to the superconducting phase is always of first order,  and is of
second order below $\Sigma_c$. Also, $\Sigma_{c}$ depends on $\eta$ and $\kappa$, and $T_{c}$ is independent of $\Sigma$. 
Physically, this is consistent in our model, since near $T_{c}$ the condensate is small and therefore the effect 
of the $\Sigma$ term on $T_{c}$ will be negligible, see eq.(\ref{generalizedterms}).

Our results have been numerical till now, and originated from the gravity dual of a strongly coupled field theory. To 
verify that these are reliable, we will calculate the free energy in the probe limit, \footnote{It is not difficult to extend this
calculation by including backreaction. However, the expressions are lengthy, and since our aim is to merely check the consistency of the results, 
we will find it sufficient to work in the probe limit.} and thus take our background as the four dimensional AdS-Schwarzschild black hole,
\begin{eqnarray}
ds^{2}=-g(r)dt^{2}+\frac{dr^{2}}{g(r)}+r^{2}(dx^{2}+dy^{2}) \nonumber
\end{eqnarray}
where, as usual, $g(r)=r^2-{r_h^3\over r} ~ ={r_h^2\over z^2}(1-z^3)$, with $r_h = 1$.
Using the AdS/CFT dictionary one can identify the Gibbs free energy of the boundary thermal state with the bulk on-shell action. Hence we first calculate the latter, and find
 \begin{eqnarray}
S_{\rm on-shell} &=& {\mu \rho \over 2}+{3\over 2}\Psi_-
\Psi_++\frac{r_{B}}{2}\Psi_-^2-{1\over 2}\int_0^1 \mathrm{d}z {\Phi(z)^2\Psi(z)^2\over z^4 g(z)}\nonumber\\
 &+&{\eta\over 2}\int_0^1 \mathrm{d}z \frac{\Phi(z)^2
\Phi'(z)^2\Psi(z)^2}{g(z)}\Big(2+3\Sigma \Psi(z)^2\Big)\nonumber\\
&-&{\eta \over 2}\int_0^1 \mathrm{d}z z^4 g(z) \Phi'(z)^2 \Psi'(z)^2 
\end{eqnarray}
where we have set $\gamma=4$, as before, and $r_B$ is a upper cut-off before the boundary $(r\rightarrow \infty)$.
\begin{figure}[t!]
\begin{minipage}[b]{0.5\linewidth}
\centering
\includegraphics[width=2.7in,height=2.3in]{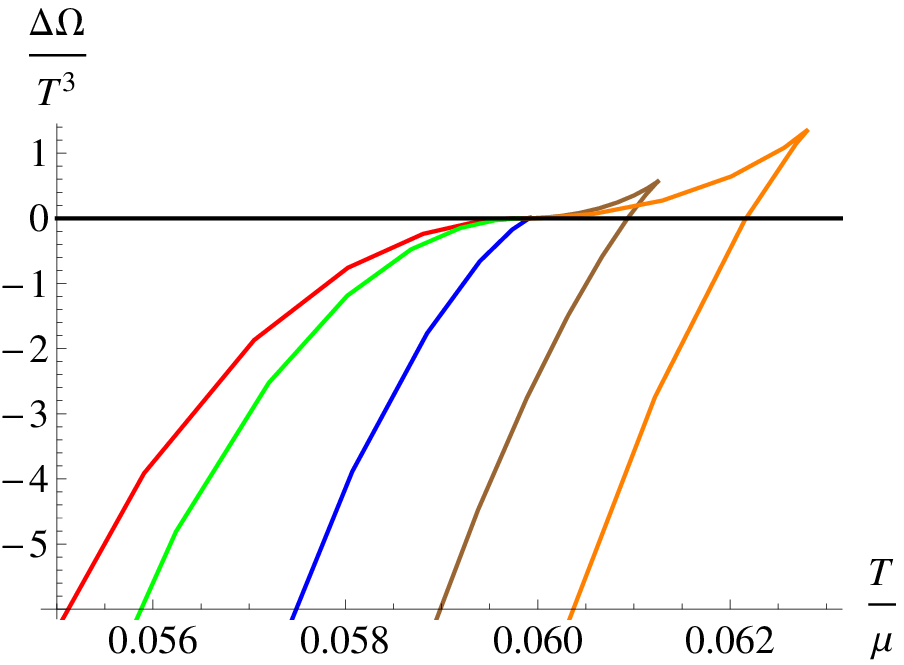}
\caption{Difference of the free energy between the superconducting and the normal phases for different values of $\Sigma$, with a fixed value of $\eta = -0.1$.
The red, green, blue, brown and orange curves correspond to $\Sigma =$ $0$, $1$, $3$, $5$ and $7$, respectively.}
\label{free1}
\end{minipage}
\hspace{0.4cm}
\begin{minipage}[b]{0.5\linewidth}
\centering
\includegraphics[width=2.7in,height=2.3in]{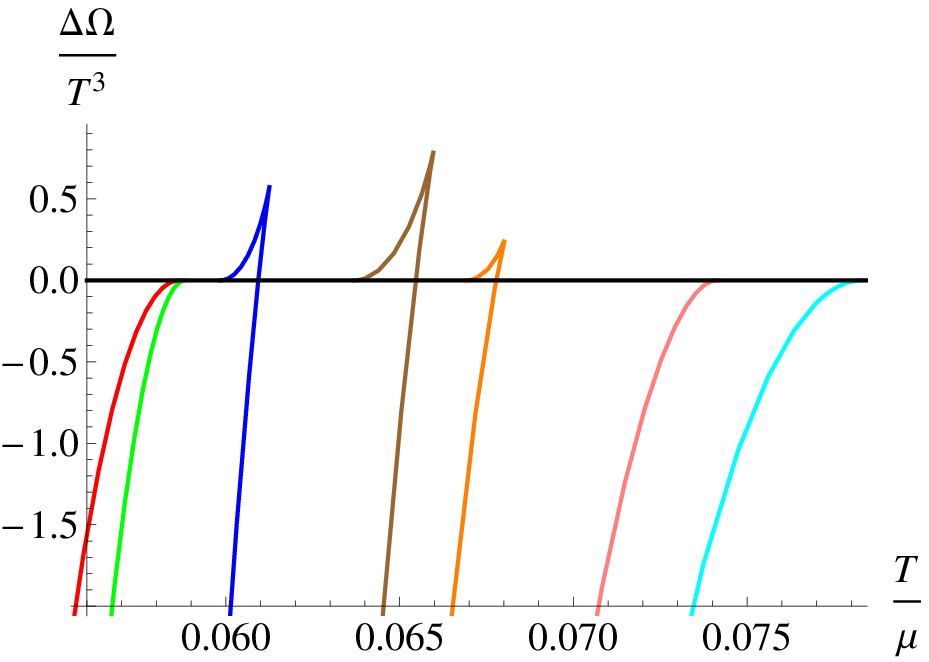}
\caption{Difference of the free energy between the superconducting and the normal phases for different values of $\eta$, with $\Sigma = 5$.
The red, green, blue, brown, orange, pink and cyan curves correspond to $\eta =$ $0.01$, $-0.01$, $-0.1$, $-0.5$, $-1$, $-3$ and $-5$, respectively. }
\label{free2}
\end{minipage}
\end{figure}
Because of the presence of $r_B$, $S_{\rm on-shell}$ diverges. To obtain a renormalized free energy, we add a boundary counter-term to the on-shell action,
\begin{eqnarray}
S_{\rm CT} &=& -{1\over 2}\int_{\rm Boundary}\Big[ \sqrt{-h}
\Psi(r)^2\Big]_{\rm Boundary}\nonumber\\
&=& -\frac{r_{B}}{2}\Psi_-^2 - \Psi_- \Psi_+\nonumber
\end{eqnarray}
where $h$ is the induced metric on the boundary. Finally the renormalized on-shell action becomes,
\begin{eqnarray}
S_{\rm renorm}=S_{\rm on-shell}+S_{\rm CT}
\label{renorm}
\end{eqnarray}
Now with the action of eq.(\ref{renorm}), we can calculate the renormalized free energy, given by $\Omega = -S_{\rm renorm}$.
The difference in the free energy between the superconducting and the normal phases can be shown to be given by \footnote{There is an extra term proportional to 
$\langle\mathcal{O}_{1}\rangle \langle\mathcal{O}_{2}\rangle$. This term goes to zero because of our 
boundary condition, i.e $\langle\mathcal{O}_{1}\rangle =0$.}
 \begin{eqnarray}
\Delta \Omega &=& \Omega_{\rm Superconductor}-\Omega_{\rm Normal}\nonumber\\
&=& -{\mu \rho \over 2}+{1\over 2}\int_0^1 \mathrm{d}z {\Phi(z)^2
\Psi(z)^2\over z^4 g(z)}+{\eta \over 2}\int_0^1 \mathrm{d}z z^4 g(z) \Phi'(z)^2 \Psi'(z)^2 \nonumber\\
&-&{\eta\over 2}\int_0^1 \mathrm{d}z \frac{\Phi(z)^2
\Phi'(z)^2\Psi(z)^2}{g(z)}\Big(2+3\Sigma \Psi(z)^2\Big) +{\mu^2\over 2}
\label{diffFreeEnergy}
\end{eqnarray}
In figs.(\ref{free1}) and (\ref{free2}), we show the variation of the difference in free energy between the superconducting and the normal phases (eq.(\ref{diffFreeEnergy})). 
In fig.(\ref{free1}), the red, green, blue, brown and orange curves correspond to $\Sigma =$ $0$, $1$, $3$, $5$ and $7$, respectively. In fig.(\ref{free2}), the red, green, blue,
brown, orange, pink, and cyan curves correspond to $\eta =$ $0.01$, $-0.01$, $-0.1$, $-0.5$, $-1$, $-3$ and $-5$, respectively. 
It should be obvious to the reader that this conforms to our earlier discussion on the existence of a window of $\eta$ in which the phase transition becomes of first order.
 
\section{\textbf{Computing the Conductivity}}

In this section we will study transport properties, mainly the optical conductivity of our boundary superconducting system. To study optical properties of the boundary system,
we introduce gauge field and metric perturbations in bulk. We will work with vector type perturbations where
\begin{eqnarray}
g_{xt}\neq 0, \ \ g_{xy}\neq 0
\end{eqnarray}
Assuming spatial and time dependence of the form $e^{i(ky-\omega t)}$, and working at the linearized level, one can show that only the $x$-component of the gauge field $A_{x}$ is relevant
to the analysis. In these perturbations, only the $(x,r)$, $(x,y)$ and $(x,t)$ components of the Einstein's  equation (eq.(\ref{einsteineom})) along with the $A_{x}$ equation are independent. 
However, in the $k\rightarrow0$ limit, which is sufficient for us to calculate the optical conductivity, it can be shown that the $(x,y)$ and the $(x,t)$ equations decouple. 
Therefore we are left only with $A_{x}$ and $(x,r)$ EOMs. Further, by substituting the $(x,r)$ equation into the equation
for $A_{x}$, we find that the $A_{x}$ equation decouples from all the gravity perturbations. After implementing the above steps, we get the $A_{x}$ EOM as
\begin{eqnarray}
&& A_{x}''\left(1+\eta  g \psi '^2\right) + A_{x}' \left(\frac{g'}{g}-\frac{\chi '}{2}+2 \eta  g' \psi'^2
-\frac{1}{2} \eta  g \chi ' \psi '^2+2 \eta  g \psi ' \psi ''\right) \nonumber \\ && + 2\kappa^{2} \eta ^2 e^{\chi } A_{x}  \left(-\frac{e^{2 \chi }  
\text{K}(\Psi)^2 \phi ^4 \phi'^2}{g^3} + \frac{2 e^{\chi } \text{K}(\Psi) \phi ^2 \phi '^2 \psi'^2}{g} -   g \phi '^2 \psi '^4\right)
\nonumber \\ && +A_{x} \left(\frac{e^{\chi } \omega^2}{g^2}-\frac{\text{G}(\Psi)}{g}-\frac{2\kappa^{2} e^{\chi }\phi '^2}{g} \right)  + \eta  e^{\chi }  
\textrm{K}(\Psi) A_{x} \biggl(-\frac{e^{\chi } \omega ^2  \phi^2}{g^3} \nonumber \\  && - \frac{\phi  \textrm{K}(\Psi)' \phi '}{g \textrm{K}(\Psi)}-
\frac{ \phi'^2}{g}+\frac{4\kappa^{2} e^{\chi}  \phi ^2 \phi'^2}{g^2}-\frac{ \phi  \phi ' \chi '}{2 g}- \frac{4 \kappa^{2}  \phi'^2 \psi '^2}{\text{K}(\Psi)}-\frac{ \phi  \phi''}{g}\biggr)=0
\label{Axeom}
\end{eqnarray}

\begin{figure}[t!]
\begin{minipage}[b]{0.5\linewidth}
\centering
\includegraphics[width=2.7in,height=2.3in]{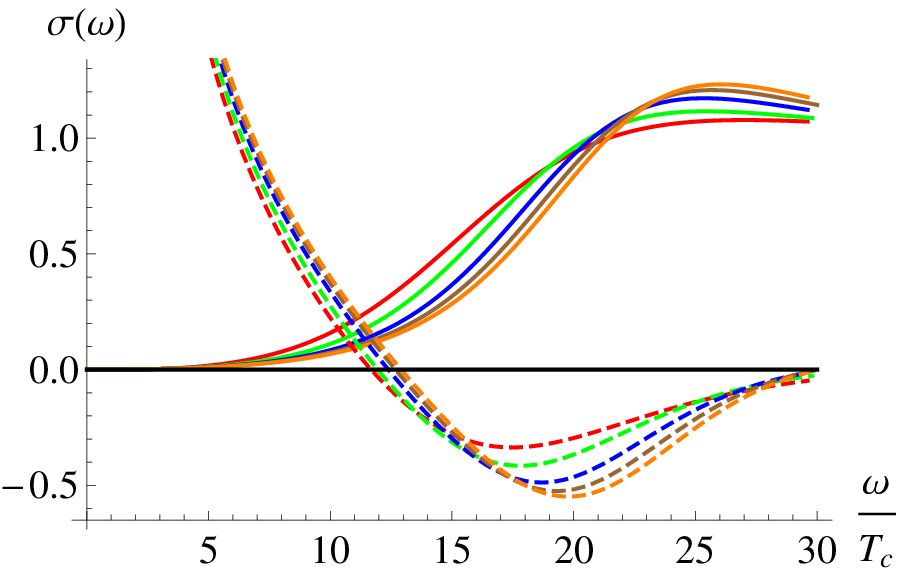}
\caption{Real (solid lines) and imaginary (dotted lines) part of conductivity for fixed $2\kappa^{2}=0.5$ and $\eta=-0.1$. The red, green, blue, brown and orange curves correspond 
to $\Sigma$= 0, 1, 3, 5 and 7, respectively.}
\label{condVsSigmaAlpha0.5Eta-0.1}
\end{minipage}
\hspace{0.4cm}
\begin{minipage}[b]{0.5\linewidth}
\centering
\includegraphics[width=2.7in,height=2.3in]{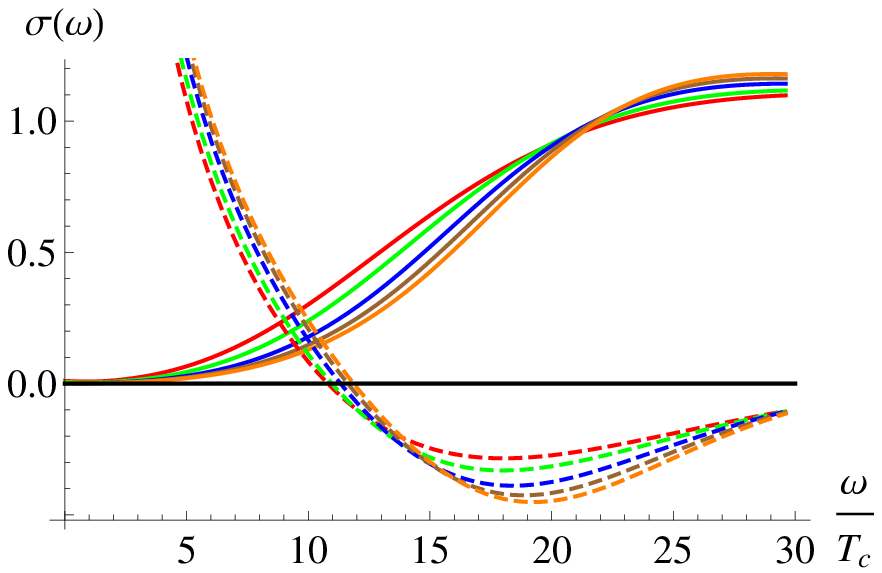}
\caption{Real (solid lines) and imaginary (dotted lines) part of conductivity for fixed $2\kappa^{2}=0.5$ and $\eta=-1$. The red, green, blue, brown and orange curves correspond 
to $\Sigma$= 0, 1, 3, 5 and 7, respectively.}
\label{condVsSigmaAlpha0.5Eta-1}
\end{minipage}
\end{figure}

\begin{figure}[h!]
\begin{minipage}[b]{0.5\linewidth}
\centering
\includegraphics[width=2.7in,height=2.3in]{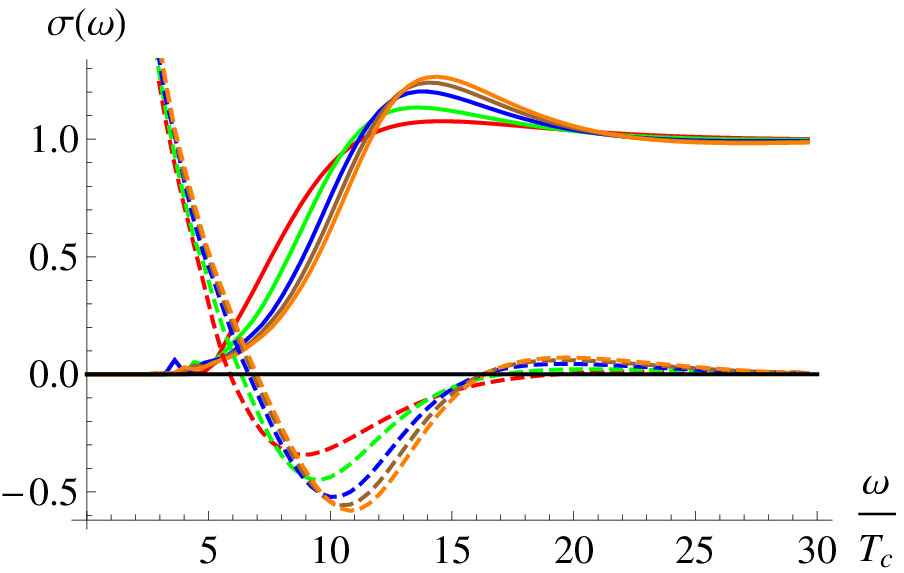}
\caption{Real (solid lines) and imaginary (dotted lines) part of conductivity for fixed $2\kappa^{2}=0$ and $\eta=-0.1$. The red, green, blue, brown and orange curves correspond 
to $\Sigma$= 0, 1, 3, 5 and 7, respectively.}
\label{condVsSigmaAlpha0Eta-0.1}
\end{minipage}
\hspace{0.4cm}
\begin{minipage}[b]{0.5\linewidth}
\centering
\includegraphics[width=2.7in,height=2.3in]{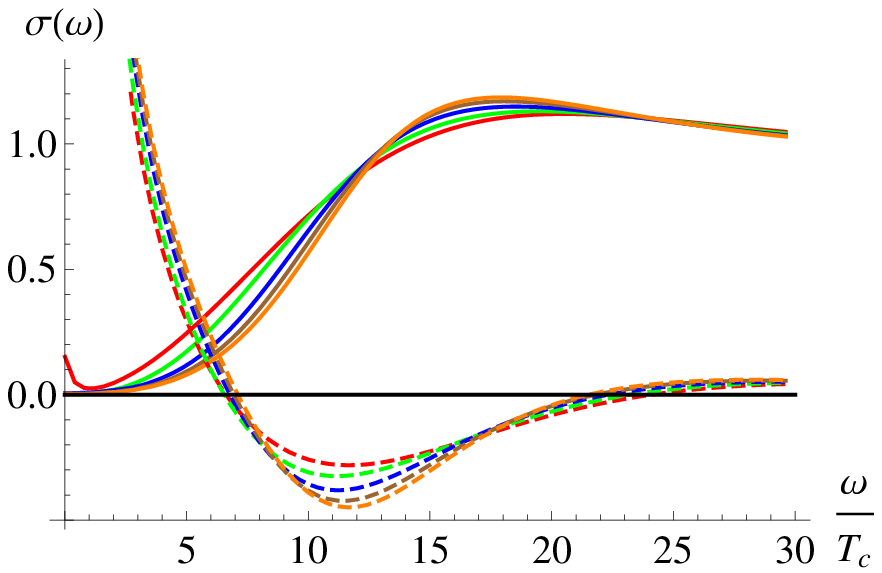}
\caption{Real (solid lines) and imaginary (dotted lines) part of conductivity for fixed $2\kappa^{2}=0$ and $\eta=-1$. The red, green, blue, brown and orange curves correspond 
to $\Sigma$= 0, 1, 3, 5 and 7, respectively.}
\label{condVsSigmaAlpha0Eta-1}
\end{minipage}
\end{figure}

We need to solve this equation with suitable boundary conditions. At the horizon, we impose infalling boundary conditions i.e $A_{x} \ \propto \  g(r)^{-i \omega/{4 \pi T_{H}}}$, where
$T_H$ is the Hawking temperature of the black hole defined in eq.(\ref{Hawking}). 
At the asymptotic boundary, $A_{x}$ behaves as
\begin{equation}
A_{x}=A_{x}^{(0)}+\frac{A_{x}^{(1)}}{r} + \cdots
\end{equation}
Using the AdS/CFT correspondence, $A_{x}^{(0)}$ and $A_{x}^{(1)}$ can be identified as the dual source and the expectation value of the boundary current, respectively. 
In order to calculate the optical conductivity, we first need the current-current correlator \cite{Robert}. 
This can be evaluated using the prescription of \cite{Son}, and the expression for the conductivity $\sigma$ can be shown to be given by
\begin{equation}
\sigma(\omega)=-\frac{i A_{x}^{(1)}}{\omega A_{x}^{(0)} }
\end{equation} 
Due to the nonlinear nature of eq.(\ref{Axeom}), we will resort to numerical analysis, and use the $z$-coordinate to perform numerical computations. 
We now discuss of the main features of the conductivity, which we present graphically.

In fig.(\ref{condVsSigmaAlpha0.5Eta-0.1}), the variation of $\sigma(\omega)$ with respect to $\omega/T_{c}$  for a fixed value of the back reaction parameter $2\kappa^{2}=0.5$ and $\eta=-0.1$ 
are shown. In figs.(\ref{condVsSigmaAlpha0.5Eta-0.1}) - (\ref{condVsSigmaAlpha0Eta-1}), the  red, green, blue, brown and orange curves correspond 
to $\Sigma$= 0, 1, 3, 5 and 7, respectively. The solid and dashed lines represents ${\rm Re}(\sigma)$ and ${\rm Im}(\sigma)$, respectively, and we have 
chosen $T=0.2 T_{c}$ \footnote{In this and subsequent figures, the temperatures are measured in appropriate units of $\rho$, the charge density of the
boundary theory.}. We find a gap in frequency $\omega_{g}$ for all the cases studied. Near this gap, the conductivity rises very gradually. The ratio of the gap frequency to 
the critical temperature $(\omega_{g}/T_{c})$, which is related to the strongly coupled nature of boundary system, increases with $\Sigma$. But importantly, we note 
that $(\omega_{g}/T_{c})$ is almost twice as large, as compared to cases studied in \cite{Robert}, where the authors found that $(\omega_{g}/T_{c})\simeq 8$ (other 
exceptions of this ratio are also known, see e.g \cite{Soda}\cite{Wang}). For different value of $\eta$, the results for the conductivity are qualitatively similar. 
This is shown in fig.(\ref{condVsSigmaAlpha0.5Eta-1}). 

However, in the probe limit, substantial differences emerge. For $\kappa^{2}=0$, the gap in 
conductivity is shifted to the left. This is shown in figs.(\ref{condVsSigmaAlpha0Eta-0.1})-(\ref{condVsSigmaAlpha0Eta-1}). 
This suggests that both $\Sigma$ and $\kappa$ play significant roles in deciding the magnitude of $(\omega_{g}/T_{c})$ in the boundary superconducting theory. In particular,
we see that backreaction effects might substantially increase the frequency gap. We note here that the behavior of the real part of the conductivity is somewhat noisy
in the lower frequency regions in fig.(\ref{condVsSigmaAlpha0Eta-0.1}) and also ${\rm Re}(\sigma)$ seems to rise for very low frequencies in fig.(\ref{condVsSigmaAlpha0Eta-1}). 
These are possible numerical artefacts of the probe limit, and are cured by 
back reaction effects as can be seen from figs.(\ref{condVsSigmaAlpha0.5Eta-0.1}) and (\ref{condVsSigmaAlpha0.5Eta-1}).

\begin{figure}[h!]
\begin{minipage}[b]{0.5\linewidth}
\centering
\includegraphics[width=2.7in,height=2.3in]{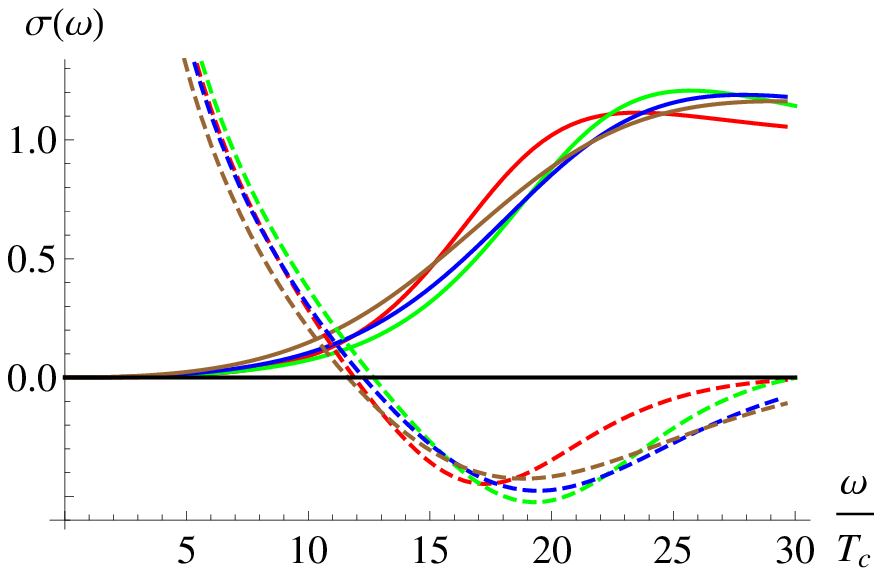}
\caption{Real (solid lines) and imaginary (dotted lines) part of conductivity for fixed $2\kappa^{2}=0.5$ and $\Sigma=5$. Red, green, blue and brown curves correspond to $\eta$= -0.01, -0.1, -0.5 and -1, respectively.}
\label{condVsEtaAlpha0.5Sigma5}
\end{minipage}
\hspace{0.4cm}
\begin{minipage}[b]{0.5\linewidth}
\centering
\includegraphics[width=2.7in,height=2.3in]{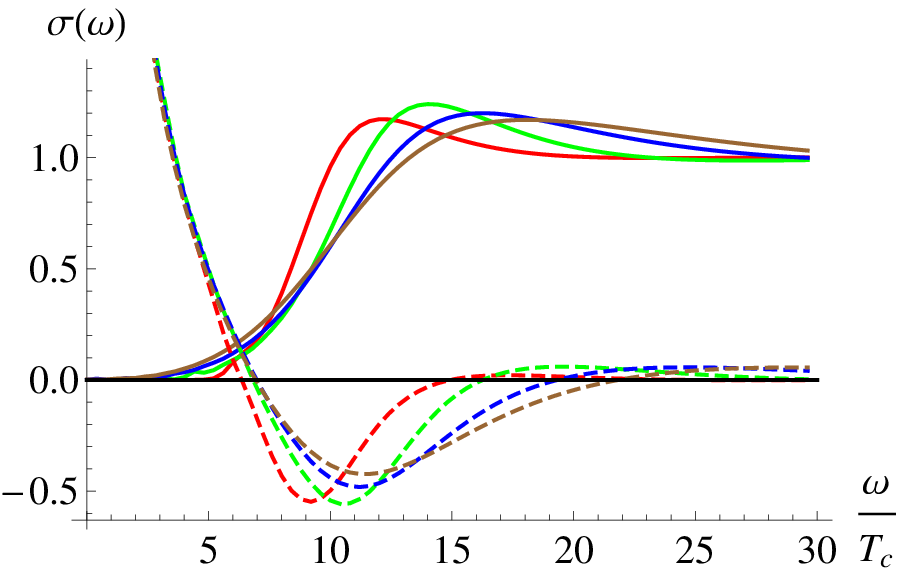}
\caption{Real (solid lines) and imaginary (dotted lines) part of conductivity for fixed $2\kappa^{2}=0$ and $\Sigma=5$. The Red, green, blue and brown curves here 
correspond to $\eta$= -0.01, -0.1, -0.5 and -1, respectively.}
\label{condVsEtaAlpha0Sigma5}
\end{minipage}
\end{figure}

For the sake of completeness, the variation of conductivity for different values of $\eta$ are shown in figs.(\ref{condVsEtaAlpha0.5Sigma5}) - (\ref{condVsEtaAlpha0Sigma5}). 
Again, the calculation is done at $T=0.2 T_{c}$.

\section{\textbf{Analytic results for the Critical Temperature}}

 In this section we will calculate the critical temperature of the boundary theory analytically.\footnote{In this section, $\eta$ and $\kappa$ will be assumed to be small. This is in contrast to the numerical
 calculations  presented in sections 2 and 3. In particular, we ignore terms of the order of $\eta^2$ and $\kappa^4$.}
 
One can notice from figs.(\ref{O2VsSigmaAlpha0.5Eta-0.1})-(\ref{O2VsSigmaAlpha0Eta-1}) 
 that for fixed values of $\eta$ and $2\kappa^{2}$, $T_{c}$ or equivalently $\mu_{c}$  is independent of $\xi$ and $\Sigma$. Therefore, in the analytic derivation of $T_{c}$, 
 we can safely put $\xi=0$ and $\Sigma=0$. In  what follows, we will use the $z$-coordinate, as before. Near $T_{c}$, when the condensate is small, one can use $\langle \textit{O}_{2} \rangle=\epsilon$  
 as an expansion parameter. Following \cite{HerzogAnalytic}-\cite{Yao}, we expand $\Psi(z)$, $\Phi(z)$, $g(z)$ and $\chi(z)$ as

 \begin{eqnarray}
\Phi &=& \Phi_{0}+\epsilon^{2}\Phi_{2}+\epsilon^{4}\Phi_{4} + \cdots,~~~ \Psi=\epsilon\Psi_{1}+\epsilon\Psi_{2}+\epsilon\Psi_{3} + \cdots \nonumber \\
g &=& g_{0}+\epsilon^{2}g_{2}+\epsilon^{4}g_{4} + \cdots,~~~\chi=\epsilon^{2}\chi_{2}+\epsilon^{4}\chi_{4} + \cdots
\label{expansion}
 \end{eqnarray}
in this formalism, the chemical potential will be corrected order by order,
 \begin{eqnarray}
\mu=\mu_{0}+\epsilon^{2}\delta\mu_{2} + \cdots
 \end{eqnarray}
with the condition that $\delta\mu_{2}$ and other higher order terms are zero at the critical point. Therefore we can identify $\mu_{0}$ as the critical chemical potential $\mu_{c}$.
Now we will calculate various quantities appearing in eq.(\ref{expansion}). At zeroth order, we have
\begin{equation}
\Phi_{0}''(z)=0,~~
\frac{1}{z}-\frac{3 r_{h}^2}{z^3 g_{0}(z)}-\frac{ g_{0}'(z)}{ g_{0}(z)}+\frac{\kappa^{2} z
\Phi_{0}'(z)^2}{2  g_{0}(z)}=0
 \end{equation}
A solution of the above equations is given by
\begin{eqnarray}
\Phi_{0}(z)=\mu_{0}(1-z)
\label{phi0}
\end{eqnarray}
\begin{eqnarray}
g_{0}(z)=\frac{r_{h}^2}{z^2}(1- z^{3})-\frac{\kappa^{2}}{2} \mu ^2 (z - z^2 )
\label{g0}
\end{eqnarray}
Notice that the first term in eq.(\ref{g0}) is the same as the coefficient of $dt^{2}$ in AdS-Schwarzschild metric.  This is an indication that the formalism based on the above 
expansion is correct and that the second term is the correction induced by backreaction.

At the first order therefore we have
\begin{eqnarray}
\Psi_{1}''(z)\left(r_{h}^2-z^4 \eta  \Phi_{0}'(z)^2\right) +
\Psi_{1}(z) \left(-\frac{m^2 r_{h}^4}{z^4 g_{0}(z)}+\frac{ r_{h}^4 \Phi_{0}(z)^2}{z^4 g_{0}(z)^2}-\frac{r_{h}^2 \eta  \Phi_{0}(z)^2 \Phi_{0}'(z)^2}{g_{0}(z)^2}\right)  &\nonumber \\ +\Psi_{1}'(z) \left(\frac{r_{h}^2
   g_{0}'(z)}{g_{0}(z)}-4 z^3 \eta  \Phi_{0}'(z)^2-\frac{z^4 \eta  g_{0}'(z)
   \Phi_{0}'(z)^2}{g_{0}(z)}-2 z^4 \eta  \Phi_{0}'(z) \Phi_{0}''(z)\right)=0
\label{psi1eom}
\end{eqnarray}
It is difficult to find an exact solution of above equation. However, using Taylor expansion, one can obtain the near horizon ($z=1$) behaviour of $\Psi_{1}(z)$ :
\begin{eqnarray}
\Psi_{1}(z)=\Psi_{1}(1)-\Psi_{1}'(1)(1-z)+\frac{1}{2}\Psi_{1}''(1)(1-z)^{2} + \cdots
\label{Taylor}
\end{eqnarray}
Also, from eq.(\ref{psi1eom}), we get
\begin{eqnarray}
\Psi_{1}'(1)=\Psi_{1}(1) \left(\frac{m^2 r_{h}^4}{g_{0}'(1) \textit{R}}\right)
\end{eqnarray}
and similarly
\begin{eqnarray}
&& \Psi_{1}''(1)=\Psi_{1}'(1) \left(\frac{m^2 r_{h}^4}{2 g_{0}'(1) \textit{R}}+\frac{4 \eta  \Phi_{0}'(1)^2}{\textit{R}}-\frac{ g_{0}''(1)}{2 g_{0}'(1) }    
+\frac{2 \eta  \Phi_{0}'(1)  \Phi_{0}''(1)}{\textit{R}} \right) \nonumber \\ && +\Psi_{1}(1) \left(-\frac{2 m^2 r_{h}^4}{g_{0}'(1) \textit{R}}-\frac{r_{h}^2 \Phi_{0}'(1)^2}{2 g_{0}'(1)^2} \right)
\end{eqnarray}
where $\textit{R}=r_{h}^2-\eta  \Phi_{0}'(1)^2$. Therefore, near the horizon, $\Psi_{1}(z)$ is given by
\begin{eqnarray}
\Psi_{1}(z)=\Psi_{1}(1) - \Psi_{1}(1) \left(\frac{m^2 r_{h}^4}{g_{0}'(1) \textit{R}}\right)(1-z)+\Psi_{1}(1) \biggl(-\frac{2 m^2 r_{h}^4}{g_{0}'(1) \textit{R}}-\frac{r_{h}^2 \Phi_{0}'(1)^2}{2 g_{0}'(1)^2}  &\nonumber \\
+\biggl(\frac{m^2 r_{h}^4}{g_{0}'(1) \textit{R}}\biggr) \biggl(\frac{m^2 r_{h}^4}{2 g_{0}'(1) \textit{R}}+\frac{4 \eta  \Phi_{0}'(1)^2}{\textit{R}}-\frac{ g_{0}''(1)}{2 g_{0}'(1) }    
+\frac{2 \eta  \Phi_{0}'(1) \Phi_{0}''(1)}{\textit{R}} \biggr)\biggr)\frac{(1-z)^{2}}{2}
\label{psi1sol}
\end{eqnarray}
Near the boundary $\Psi_{1}$ falls as
\begin{eqnarray}
\Psi_{1}(z)\sim \textit{O}_{+}z^{\lambda_{+}}
\label{psi1boundary}
\end{eqnarray}
In order to obtain $T_{c}$, we will use the matching method technique of \cite{Soda} by equating eqs.(\ref{psi1sol}) and (\ref{psi1boundary}) at some intermediate point, say $z=z_{m}$.
This yields the following equations
\begin{eqnarray}
\textit{O}_{+}z_{m}^{\lambda_{+}}=\Psi_{1}(1) - \Psi_{1}(1) \left(\frac{m^2 r_{h}^4}{g_{0}'(1) \textit{R}}\right)(1-z_{m})+\Psi_{1}(1) \biggl(-\frac{2 m^2 r_{h}^4}{g_{0}'(1) 
\textit{R}}-\frac{r_{h}^2 \Phi_{0}'(1)^2}{2 g_{0}'(1)^2}  &\nonumber \\
+\biggl(\frac{m^2 r_{h}^4}{g_{0}'(1) \textit{R}}\biggr) \biggl(\frac{m^2 r_{h}^4}{2 g_{0}'(1) \textit{R}}+\frac{4 \eta  \Phi_{0}'(1)^2}{\textit{R}}-\frac{ g_{0}''(1)}{2 g_{0}'(1) }    
+\frac{2 \eta  \Phi_{0}'(1) \Phi_{0}''(1)}{\textit{R}} \biggr)\biggr)\frac{(1-z_{m})^{2}}{2}
\label{matchsol1}
\end{eqnarray}
and
\begin{eqnarray}
\lambda_{+} \textit{O}_{+}z_{m}^{\lambda_{+}-1}= \Psi_{1}(1) \left(\frac{m^2 r_{h}^4}{g_{0}'(1) \textit{R}}\right)-\Psi_{1}(1) \biggl(-\frac{2 m^2 r_{h}^4}{g_{0}'(1) \textit{R}}-
\frac{r_{h}^2 \Phi_{0}'(1)^2}{2 g_{0}'(1)^2}  &\nonumber \\
+\biggl(\frac{m^2 r_{h}^4}{g_{0}'(1) \textit{R}}\biggr) \biggl(\frac{m^2 r_{h}^4}{2 g_{0}'(1) \textit{R}}+\frac{4 \eta  \Phi_{0}'(1)^2}{\textit{R}}-\frac{ g_{0}''(1)}{2 g_{0}'(1) }    
+\frac{2 \eta  \Phi_{0}'(1) \Phi_{0}''(1)}{\textit{R}} \biggr)\biggr)(1-z_{m})
\label{matchsol2}
\end{eqnarray}
From eqs.(\ref{matchsol1}) and (\ref{matchsol2}), we get
\begin{eqnarray}
 \textit{O}_{+}= \frac{2 z_{m}^{1-\lambda_{+}}\Psi_{1}(1)}{\lambda_{+}(1-z_{m})+2z_{m}} \left[1-\frac{m^2 r_{h}^4}{g_{0}'(1) \textit{R}}\frac{(1-z_{m})}{2}\right]
\end{eqnarray}
using eq.(\ref{matchsol2}) and substituting the form of $\Phi_{0}(1)$ and $g_{0}(1)$ from eq.(\ref{phi0}) and (\ref{g0}), we get a quartic equation for $\mu_{0}$
(ignoring terms of order $\kappa^4$ and $\eta^2$) 
\begin{eqnarray}
M \mu_{0}^{4}+ r_{h}^{2} N \mu_{0}^{2} + r_{h}^{4} P=0
\label{mu0}
\end{eqnarray}
where we have defined the quantities
\begin{align*}
& M=2\kappa^{2}\eta \left(24e +m^2+ 14 f m^2 +2 e f m^2 \right) +8f\eta &\nonumber \\ & P=72 e+4 m^2 \left(3+8 f+8 e f+f m^2\right)  &\nonumber \\
& N= 4 f + 24\kappa^{2} e + 2\kappa^{2} m^2 \left( 1 + 6 f  + 2 e f \right)  +12 \eta m^2 \left(1 +10 f +2 e f \right)+144e\eta &\nonumber \\
& f=\frac{(1-z_{m})}{2}, \ \ e=\frac{\lambda_{+}}{(\lambda_{+}(1-z_{m})+2z_{m})}
\end{align*}
Now solving eq.(\ref{mu0}), we get
\begin{equation}
\mu_{0}^{2}=r_{h}^{2}\left[\frac{N\pm\sqrt{N^{2} - 4 M P}}{2 M}\right]
\end{equation}
Using the relation $\rho=\mu_0 r_h$ and indentifying the Hawking temperature as the temperature of the boundary theory, we arrive at the critical temperature
\begin{eqnarray}
T_c=\frac{3}{4\pi}\frac{\rho^{1/2}}{\left(\frac{N\pm \sqrt{N^2-4MP}}{2M}\right)^{1/4}}\left[1-\frac{2\kappa^2}{12}\left(\frac{N\pm \sqrt{N^2-4MP}}{2M}\right)\right]
\end{eqnarray}

In Table (\ref{table1}), we show a comparison between the numerical and analytical results for the critical temperature.
\begin{table}[ht]
\begin{center}
\begin{tabular}{|c|c|c|c|c|c|c|}
\multicolumn{7}{c}
{ Analytical Values of $\frac{T_c}{\sqrt{\rho}}$} \\
\hline
$2\kappa^2$ $\Big\backslash \eta$ &-0.5 &-0.3 &-0.1 &-0.01&0&0.01\\
\hline
$0$    &0.1226&0.1221&0.1204&0.1152&0.1129&0.1046\\
\hline
$0.001$  &0.1224&0.1220&0.1203&0.1152&0.1128&0.1049\\
\hline
$0.01$  &0.1212&0.1207&0.1191&0.1144&0.1123&0.1065\\
\hline
\multicolumn{7}{c}{ Numerical Values of of $\frac{T_c}{\sqrt{\rho}}$} \\
\hline
$2\kappa^2$ $\Big\backslash \eta$ &-0.5 &-0.3 &-0.1 &-0.01&0&0.01\\
\hline
$0$   &0.1233&0.1216&0.1195&0.1185&0.1184&0.1183\\
\hline
$0.001$  &0.1232&0.1215&0.1194&0.1184&0.1183&0.1182\\
\hline
$0.01$  &0.1220&0.1203&0.1181&0.1171&0.1170&0.1169\\
\hline
\end{tabular}
\caption{
Analytical and numerical values of $\frac{T_c}{\sqrt{\rho}}$ for different
values of the higher derivative coupling constant $\eta$ and the backreaction
parameter $\kappa$ at
$z_m =0.4$.}
\label{table1}
\end{center}
\end{table}
It can be checked that if we expand $\Psi_1(z)$ in eq.(\ref{Taylor}) upto third order, the change in the numerical value is only at the third decimal place. Hence, our analytic method of expanding
$\Psi_1(z)$ upto second order is trustable. \footnote{For some values of $\kappa^2$ and $\eta$, the analytical and numerical results for the critical temperature differ at the second decimal place.
This means that in principle, one should retain corrections beyond the third order in eq.(\ref{Taylor}) to compare the numerical values with the analytic ones. In this paper, we have not performed
this analysis, and our results in Table (1) is restricted to the second order expansion in eq.(\ref{Taylor}).}

\section{\textbf{Response Functions and Optical Properties}}
In this section we will briefly discuss the response functions of generalized holographic superconductors with higher derivative couplings, by considering the electromagnetic perturbation
(for more details see \cite{policastro} \cite{amaritihydro}). \footnote{We remind the reader that as pointed out in the introduction, there are no dynamical photons at the boundary. We assume 
our system to be weakly coupled to such a dynamical photon. This is standard in the literature.}
For simplicity, we will work in the probe limit. The calculations including backreaction are complicated and involve a large number of differential equations which are difficult to solve even numerically and 
we postpone this study to a future publication. Here, we will follow the conventions used in \cite{Gao},\cite{mps}, and calculate the response functions that are given in terms of the electric permittivity $\epsilon$
and the magnetic permeability $\mu$. In our holographic set up, these response functions are 
\begin{equation}
\epsilon \left( \omega \right) = 1+\frac{4\pi }{\omega ^{2}} G_{T}^{0}\left( \omega \right)
\label{eps}
\end{equation}
\begin{equation}
\mu\left( \omega \right) = \frac{1}{1- 4\pi G_{T}^{2}\left( \omega \right) }
\label{mu}
\end{equation}
where $G_{T}^{0}$ and $G_{T}^{2}$ are the coefficients of powers of the spatial momentum $k$, in the series expansion of the transverse current-current correlator $G_{T}$,
\begin{equation}
G_{T}\left( \omega,k \right) =G_{T}^{0}\left( \omega \right) +k^{2} G_{T}^{2}\left( \omega \right) + \cdots
\label{GT}
\end{equation}
Here we take finite $k$ to calculate the response functions, which is different from the previous section where we used $k=0$ for the conductivity calculation. 
By taking the same series expansion of $A_{x}$ as in $G_{T}$, we arrive the following equations for $G_{T}^{0}$ and $G_{T}^{2}$
\begin{equation}
G_{T}^0 = \frac{A_{x0}^{(1)}}{A_{x0}^{(0)}},~~~~
G_{T}^2 = \frac{A_{x0}^{(1)}}{A_{x0}^{(0)}} \biggl(\frac{A_{x2}^{(1)}}{A_{x0}^{(1)}} - \frac{A_{x2}^{(0)}}{A_{x0}^{(0)}} \biggr)
\label{GT0}
\end{equation}
Generally, the existence of negative refractive index in our strongly coupled medium can be  predicted by using the Depine-Lakhtakia (DL) index $n_{DL}$ \cite{Depine} given by
\begin{equation}
n_{DL}= Re(\epsilon )| \mu | +Re(\mu)| \epsilon  |
\label{deplak}
\end{equation}
If the DL index is negative, this indicates that the phase velocity in the medium and the direction of energy flow are opposite
to each other, and hence the system behaves like a meta material, i.e has a negative index of refraction. There is a caveat here, namely that the magnetic permeability appearing in 
eq.(\ref{mu}) is an effective permeability arising from the $\epsilon-\mu$ approach, and not the real permeability. However, as is standard in the literature \cite{policastro}, we will define 
the DL index in terms of the effective permeability. There are a few subtleties associated with the DL index, and at this point, it
is instructive to take a brief digression regarding the physics of eq.(\ref{deplak}). 

For the moment, let us concentrate on real materials. \footnote{After we completed a revised version of this paper, we learnt that the following analysis for real
materials has substantial overlap with the results of \cite{kinsler}.} For such a material, let us consider a plane electromagnetic wave that propagates along the $z$ direction. 
The Poynting vector ${\vec P}$ is parallel to the $z$ axis, and its time average is given by a standard textbook result,
\begin{equation}
P(n) = {\rm Re}\left(\frac{n}{\mu}\right)\frac{|A|^2}{2\eta}{\rm exp}~\left(-2k_0{\rm Im}(n)z\right)
\label{Poynting}
\end{equation}
where $\eta$ is the free space impedance. 
Hence, the sign of $P(n)$ is given by that of the real part of $n/\mu$, where $n = \pm\sqrt{\epsilon\mu}$ is the refractive index, $\epsilon$ being the electric permittivity and $\mu$ 
the magnetic permeability, respectively. This means that the sign of ${\rm Re}(n/\mu)$ indicates the direction of power flow, while that of ${\rm Re}(n)$ determines the direction of the phase velocity. 
In metamaterials, these two signs are expected to be opposite. However, there are a few subtleties here which are worth noting, and will be relevant for our discussion on holographic
metamaterials. 

Let us first briefly recapitulate the analysis by Depine and Lakhtakia \cite{Depine}. We write the complex valued quantities 
\begin{equation}
\epsilon = |\epsilon|~{\rm exp}(i\phi_{\epsilon}),~~\mu = |\mu|~{\rm exp}(i\phi_{\mu})
\label{epsamu}
\end{equation}
DL assume that ${\rm Im}(\epsilon)$ and ${\rm Im}(\mu)$ are positive definite and this necessarily implies that 
\begin{equation}
0\leq \phi_{\epsilon, \mu} \leq \pi
\label{angle}
\end{equation}
The complex refractive index is given by 
\begin{equation}
n = \pm \sqrt{|\epsilon|~|\mu|}~{\rm exp}~i\left(\frac{\phi_{\epsilon} + \phi_{\mu}}{2}\right)
\end{equation}
Hence we end up with the following conditions : 
\begin{eqnarray}
{\rm Re} (n) &=& \pm\sqrt{|\epsilon|~|\mu|}~{\rm cos}\left(\frac{\phi_{\epsilon} + \phi_{\mu}}{2}\right)\nonumber\\
{\rm Re}\left(\frac{n}{\mu}\right) &=& {\rm Re}\left[\pm\sqrt{\frac{|\epsilon|}{|\mu|}}~{\rm exp}~i\left(\frac{\phi_{\epsilon} - \phi_{\mu}}{2}\right)\right] = \pm\sqrt{\frac{|\epsilon|}{|\mu|}}~
{\rm cos}\left(\frac{\phi_{\epsilon} - \phi_{\mu}}{2}\right)
\label{cond1}
\end{eqnarray}
The condition that the energy flow is directed opposite to the phase velocity boils down to the following condition between the angles 
\begin{equation}
\frac{{\rm cos}\frac{1}{2}\left(\phi_{\epsilon} + \phi_{\mu}\right)}{{\rm cos}\frac{1}{2}\left(\phi_{\epsilon} - \phi_{\mu}\right)} < 0
\label{val1}
\end{equation}
\begin{figure}[h!]
\begin{minipage}[b]{0.5\linewidth}
\centering
\includegraphics[width=2.7in,height=2.3in]{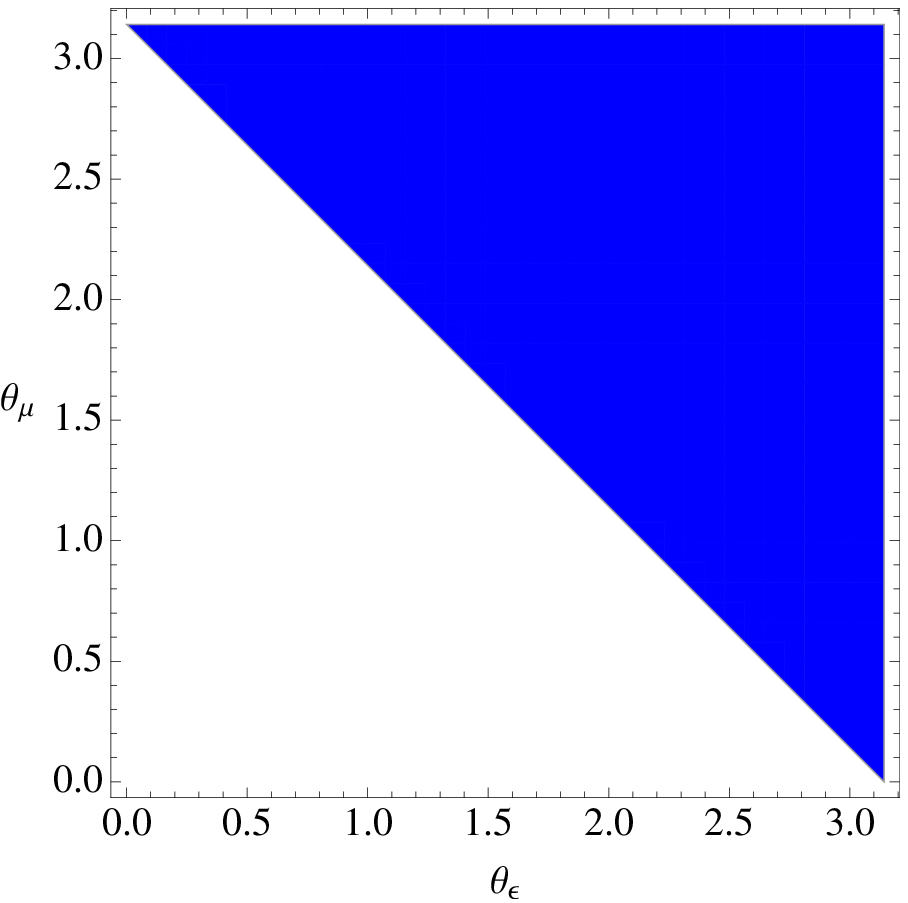}
\caption{Region of validity of the DL condition with eq.(\ref{angle}).}
\label{ndl1}
\end{minipage}
\hspace{0.4cm}
\begin{minipage}[b]{0.5\linewidth}
\includegraphics[width=2.7in,height=2.3in]{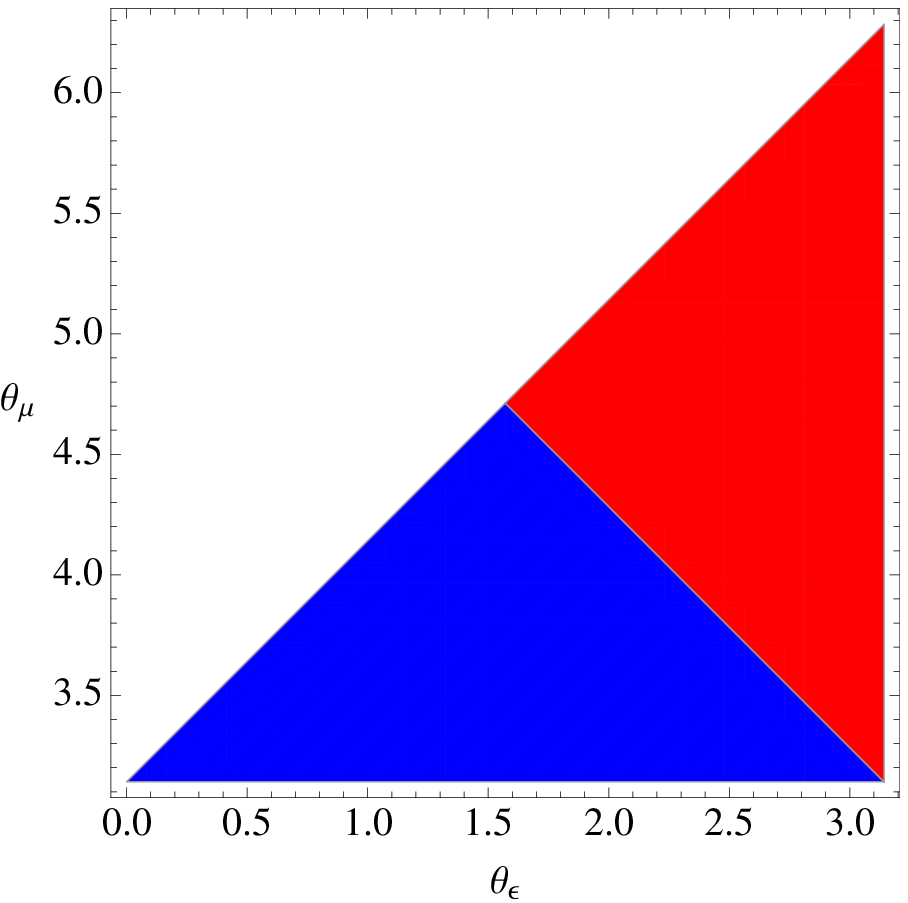}
\caption{Region of validity of the DL condition with eq.(\ref{angle2}).}
\label{ndl2}
\end{minipage}
\end{figure}
This can be seen to be satisfied for all regions in the $\phi_{\epsilon} - \phi_{\mu}$ plane for which $\phi_{\epsilon} + \phi_{\mu} \geq \pi$, as shown in fig.(\ref{ndl1}), where the region marked
in blue indicates the region of validity of eq.(\ref{val1}). It can be checked that this is (as expected) equivalent to the Depine-Lakhtakia condition 
\begin{equation}
{\rm Re}(\epsilon)|\mu| + {\rm Re}(\mu)|\epsilon| < 0 \implies |\epsilon||\mu|\left({\rm cos}\theta_{\epsilon} + {\rm cos}\theta_{\mu}\right) < 0
\label{cond2}
\end{equation}
Now the fact that the amplitude decays exponentially along the direction of the flow implies that 
\begin{equation}
\frac{{\rm Re}(n/\mu)}{{\rm Im}(n)} > 0 \implies \frac{{\rm cos}\frac{1}{2}\left(\phi_{\epsilon} - \phi_{\mu}\right)}{{\rm sin}\frac{1}{2}\left(\phi_{\epsilon} + \phi_{\mu}\right)} > 0
\label{cond3}
\end{equation}
This also means that 
\begin{equation}
{\rm Re}(n)/{\rm Im}(n) = {\rm cot} \left(\frac{\phi_{\epsilon} + \phi_{\mu}}{2}\right) < 0 
\label{rbyi}
\end{equation}
It is seen that eqs.(\ref{cond3}) and (\ref{rbyi}) are always satisfied for the region indicated in fig.(\ref{ndl1}). In an {\bf unstable} solution, where the amplitude exponentially increases
with distance, the inequalities in eqs.(\ref{cond3}) and (\ref{rbyi}) will flip. Hence, if such a condition is even a mathematical possibility, one has to look for a solution to the inequalities
\begin{equation}
\frac{{\rm cos}\frac{1}{2}\left(\phi_{\epsilon} + \phi_{\mu}\right)}{{\rm cos}\frac{1}{2}\left(\phi_{\epsilon} - \phi_{\mu}\right)} < 0,~
\frac{{\rm cos}\frac{1}{2}\left(\phi_{\epsilon} - \phi_{\mu}\right)}{{\rm sin}\frac{1}{2}\left(\phi_{\epsilon} + \phi_{\mu}\right)} < 0,~
{\rm cot} \left(\frac{\phi_{\epsilon} + \phi_{\mu}}{2}\right) > 0 
\label{fullconds}
\end{equation} 
within the angle range $0\leq \phi_{\epsilon, \mu} \leq \pi$. That this yields a null set is easily checked. Thus within the range of angles of eq.(\ref{angle}), no 
unstable solution exhibiting negative refraction is allowed. 

Now let us relax the condition on ${\rm Im}(\mu)$, and assume that it can take negative values as well (we keep ${\rm Im}(\epsilon)$ to be positive throughout this discussion, as this is what naturally 
occurs in our holographic setup). Such a scenario has been debated in the optics community and was investigated by Markel
in \cite{markel}. First note that if ${\rm Im}(\mu) < 0$, then the conditions on the angles of eq.(\ref{angle}) change, and are now given as
\begin{equation}
0 \leq \phi_{\epsilon} \leq \pi,~~~\pi \leq \phi_{\mu} \leq 2\pi
\label{angle2}
\end{equation}
If we demand that eqs.(\ref{val1}), (\ref{cond3}) and (\ref{rbyi}) be simultaneously satisfied in the range of angles of eq.(\ref{angle2}), then we have the region of validity shown in blue in fig.(\ref{ndl2}). 
However, now there is a mathematical possibility of a solution where the direction of the power flow is opposite to the phase velocity, but the amplitude grows in the direction of propagation, 
i.e the conditions of eq.(\ref{fullconds}) is satisfied, and this is depicted as the red region in fig.(\ref{ndl2}). In this region, ${\rm Re}(n)/{\rm Im}(n) > 0$. 
\begin{figure}[h!]
\centering
\includegraphics[width=2.7in,height=2.3in]{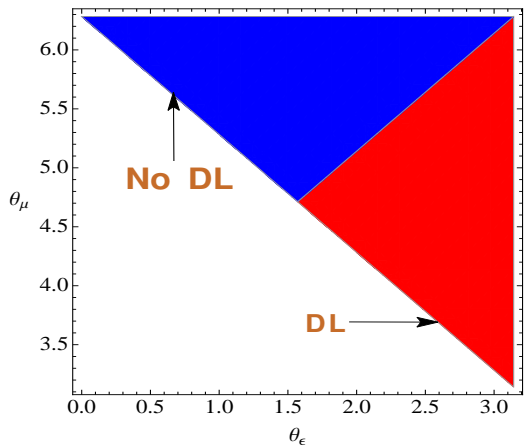}
\caption{Region of validity of the conditions of \cite{markel}. See text for details.}
\label{ndl3}
\end{figure}

However, there is an alternative interpretation of the physics of eq.(\ref{angle2}). 
In \cite{markel}, Markel proposed that for diamagnetic materials, \footnote{This is relevant as we consider holographic superconductors. However, in our setup, the boundary theory
is $2+1$ dimensional, and there are subtleties regarding expulsion of external magnetic fields in this case, as pointed out in \cite{Hartnollcorev}. We will proceed keeping this in mind.}
${\rm Im}(\mu)$ can in fact be negative. He calculated the rate of dissipation of energy from an electromagnetic wave in 
such a medium,  and showed that the Poynting vector should be proportional to ${\vec E} \times {\vec B}$ (instead of ${\vec E} \times {\vec H}$ as used conventionally). In that case, it is not hard to
see that the direction of power flow is always the same as that of the phase velocity (as the factor of $\mu$ does not occur in eq.(\ref{Poynting})). Thus, it was claimed in \cite{markel},\cite{markel1}
that for such polarizable media, negative refraction is {\it impossible}. Several claims and counter claims appeared in the optics literature after this, a full discussion of which is beyond the scope and purpose of this paper. 

We can however, make the following statement. Suppose ${\rm Im}(\mu) < 0$. Then, if ${\vec B}$ is used in the definition of the Poynting vector as advocated in \cite{markel}, then there is 
indeed no negative refraction and this, along with the fact that ${\rm Im}(\epsilon\mu) > 0$, required for the net absorption rate of heat in the medium to be positive (eq.(14) of \cite{markel}), 
can be shown to imply that the entire colored region depicted in fig.(\ref{ndl3}) is allowed. But now, if we {\it naively} apply the DL criterion (not really meaningful in this case as 
we have used ${\vec P} \propto {\vec E} \times {\vec B}$)
we do obtain a region in which the DL index is negative. This is the red region of fig.(\ref{ndl3}), marked as ``DL''. In the blue region of fig.(\ref{ndl3}), marked as ``No DL,'' the 
Depine-Lakhtakia condition does not hold. Hence we see that negativity of the DL index may not be a useful criterion even in a dissipative
medium, if the arguments of \cite{markel} hold. Specifically, the common red region of fig.(\ref{ndl2}) or fig.(\ref{ndl3}) can have very different physical interpretations. 
This rather simple result which to the best of our knowledge has not appeared in the optics literature should possibly have some experimental relevance.
\begin{figure}[h!]
\begin{minipage}[b]{0.5\linewidth}
\centering
\includegraphics[width=2.7in,height=2.3in]{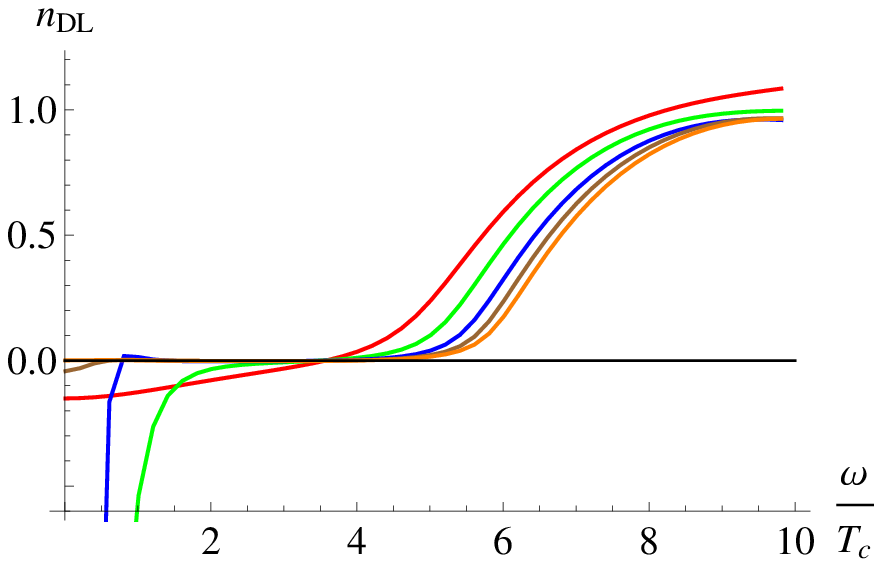}
\caption{ $n_{DL}$ with $\eta=-0.3$, $2\kappa^{2}=0$ for different values of $\Sigma$.}
\label{NDLVsSigmaAlpha0Eta-0.3}
\end{minipage}
\hspace{0.2cm}
\begin{minipage}[b]{0.5\linewidth}
\centering
\includegraphics[width=2.7in,height=2.3in]{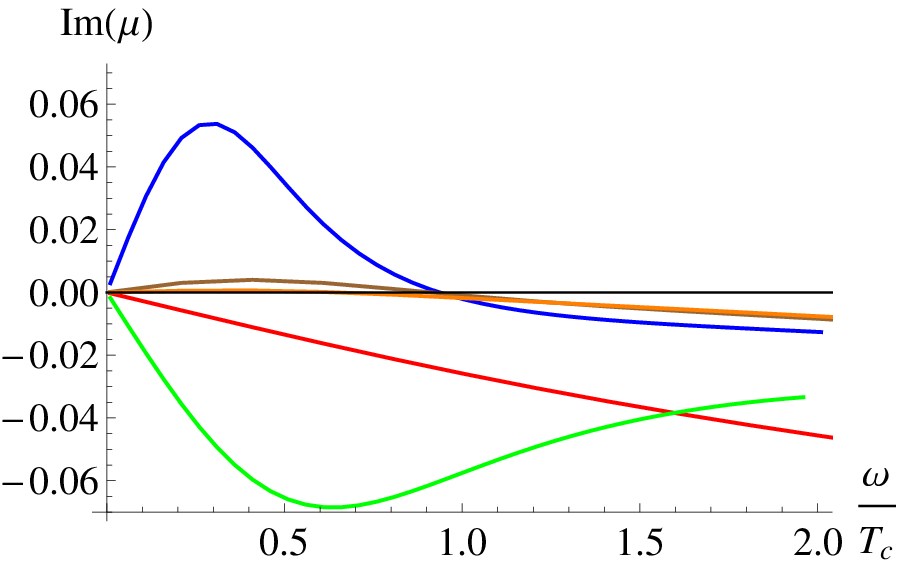}
\caption{ Im$\mu(\omega)$ with $\eta=-0.3$, $2\kappa^{2}=0$ for different values of $\Sigma$.}
\label{ImMuVsSigmaAlpha0Eta-0.3}
\end{minipage}
\end{figure}

\begin{figure}[t!]
\begin{minipage}[b]{0.5\linewidth}
\centering
\includegraphics[width=2.7in,height=2.3in]{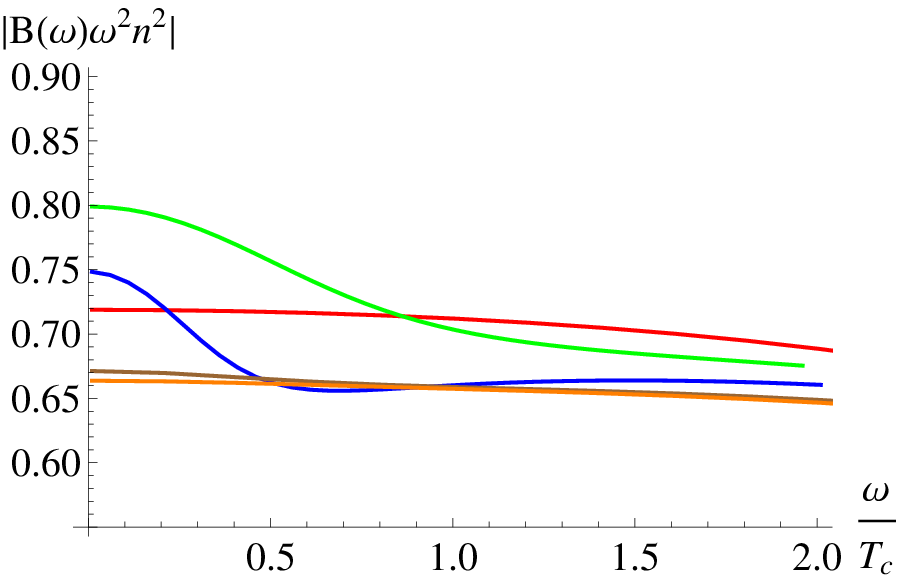}
\caption{$B(\omega)\omega^{2}n^{2}$ (see text for definition) with $\eta=-0.3$, $2\kappa^{2}=0$ for different values of $\Sigma$.}
\label{ValidEXPVsSigmaAlpha0Eta-0.3}
\end{minipage}
\hspace{0.2cm}
\begin{minipage}[b]{0.5\linewidth}
\centering
\includegraphics[width=2.7in,height=2.3in]{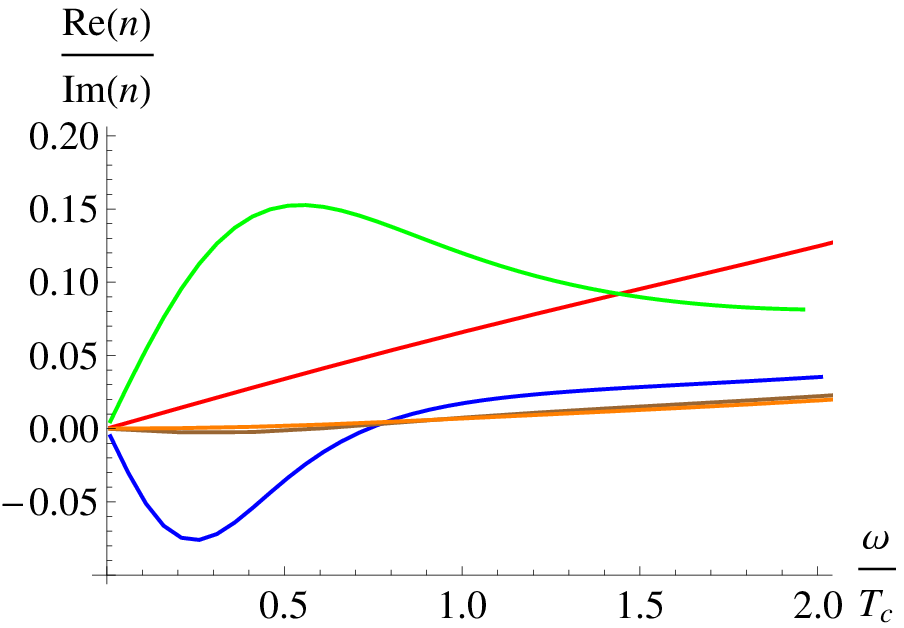}
\caption{Propagation to dissipation ratio with $\eta = -0.3$, $2\kappa^{2}=0$ for different values of $\Sigma$.}
\label{renbyimn}
\end{minipage}
\end{figure}

Now let us come back to our holographic scenario. First we present the results on the DL index. 
In fig.(\ref{NDLVsSigmaAlpha0Eta-0.3}), \footnote{The color coding for figs.(\ref{NDLVsSigmaAlpha0Eta-0.3}) - (\ref{renbyimn}) is as follows. 
The red, green, blue, brown and orange curves correspond to $\Sigma = 0, 1, 3, 5$ and $7$, respectively.}
we have shown numerical results on the variation of $n_{DL}$ with frequency, where we have taken $T= 0.5T_{c}$. 
At high frequencies, the behavior of $n_{DL}$ is qualitatively similar to the ones reported in \cite{Gao},\cite{amariti}, but at low frequencies, significant differences emerge, where we find 
that $n_{DL}$ can be negative. In R-charged black hole backgrounds, similar results in the probe limit were found in \cite{mps}. This low frequency behavior of $n_{DL}$ in the probe 
limit in the $AdS_{4}$ black-hole background should be contrasted with the results in \cite{Gao},\cite{amariti} where no negative DL index was found at any frequencies in the probe limit in 
$AdS_{5}$ backgrounds. Specifically, in the approach of \cite{amariti}, the bulk fields are expanded with respect to the order parameter near to criticality, and 
analytic results can be established very close to $T_c$, indicating this result. This approach is difficult to implement in four dimensional backgrounds, but
we do find numerically that close to criticality, the DL index is indeed positive, as in five dimensional backgrounds. Only at lower values of the temperature is the DL index negative. We do not 
have a complete understanding of the issue, but the appearance of a negative DL index in the probe limit in our four dimensional background is probably due to the different
nature of the bulk theory. We are unable to comment on this further. 

In our holographic approach, \footnote{Upon setting $\eta = 0$, i.e removing the higher derivative correction in eq.(\ref{action}), we have checked that the results of this section go over to the 
corresponding cases studied in \cite{mps}, as they should.}  we need to be careful about the interpretation of fig.(\ref{NDLVsSigmaAlpha0Eta-0.3}). For this, one should first check the sign of ${\rm Im}(\mu)$. As we have
mentioned,  $\mu$ is an effective permeability unlike the real magnetic permeability of materials discussed earlier in this section, and may not be an observable. We proceed keeping this in mind. 
From fig.(\ref{ImMuVsSigmaAlpha0Eta-0.3}), we see that the sign of  ${\rm Im}(\mu)$ is dependent on the choice of the parameter $\Sigma$. 
Specifically, the red and the green lines of fig.(\ref{ImMuVsSigmaAlpha0Eta-0.3}), corresponding to $\Sigma = 0,~1$ show negative values of ${\rm Im}(\mu)$ for small frequencies, but the situation 
changes when the value of $\Sigma$ is enhanced, and higher values of $\Sigma$ seems to push ${\rm Im}(\mu)$ to positive values. \footnote{It is well known that back reaction effects have similar properties, 
i.e generically in the probe limit, ${\rm Im}(\mu)$ is negative, while inclusion of back reaction can make this positive. } 

From our previous discussion, we see that when ${\rm Im}(\mu) < 0$, we can either interpret the negative DL index for the red and the green lines as lying in the unstable
(red) range of fig.(\ref{ndl2}), or we could interpret this as a physical solution lying in the red colored region of fig.(\ref{ndl3}), where the system is dissipative, and there is no negative refraction although
$n_{DL}$ is negative. Both interpretations look plausible and we have not been able to settle this issue. We also note that for $\Sigma \geq 3$ (the blue, brown and orange curves of 
figs.(\ref{NDLVsSigmaAlpha0Eta-0.3}) and (\ref{ImMuVsSigmaAlpha0Eta-0.3})), our strongly coupled system behaves like a real metamaterial, with ${\rm Im}(\mu) > 0$ and here the negativity of
the DL index has the normal interpretation of negative refraction. 

As a final comment here we add that in the low frequency 
region, ${\rm Im}(\mu)$ gradually increases from negative to positive values as we increase $\Sigma$ from $1$ to $3$. However, the validity of the expansion used in eq.(\ref{GT}) do not seem to be satisfied 
for the window $1.5<\Sigma<2.5$, which seems to be an unfortunate feature of the probe analysis. 
Hence we have restricted our attention to $\Sigma \leq1$ and $\Sigma \geq 3$. Of course, as we increase the value of $\Sigma$, back reaction effects cannot be completely 
neglected, and hence our probe analysis becomes less reliable. We will postpone a full analysis of response functions including back reactions to a future study.
For the values of $\Sigma$ that we have used, the validity condition of eq.(\ref{GT}) 
is shown in fig.(\ref{ValidEXPVsSigmaAlpha0Eta-0.3}). Strictly speaking, the $\epsilon-\mu$ expansion is valid when $|B n^2\omega^2| \ll 1$, with 
$G_{T}^{2}/G_{T}^{0}=B$ and $k^{2}=\omega^{2}|n|^{2}$. In our case, as 
seen from fig.(\ref{ValidEXPVsSigmaAlpha0Eta-0.3}), this is only marginally satisfied in the frequency region 
where $n_{DL}$ is negative. This caveat may be related to the appearance of a negative imaginary part of the magnetic permeability, as was pointed out in \cite{mps}. 
 
One also needs to be sure that the system is in thermodynamic equilibrium. This can be done by showing ${\rm Im}(G_{T})>0$ for real $\omega$ and $k$. 
We find that for small $k$, which is necessary for the expansion in eq.(\ref{GT}) to be valid, ${\rm Im}(G_{T})>0$ for all cases, thus ensuring 
that our system is indeed in thermodynamic equilibrium.
Finally, we point out that as in most cases of holographic optics studied so far, the propagation to dissipation ratio, given by ${\rm Re}(n)/{\rm Im}(n)$ is typically very small in our case. 
This is depicted in fig.(\ref{renbyimn}). However, we note that this ratio seems to be enhanced for some values of $\Sigma$, as compared to the others. Also, beyond $\Sigma \sim 3$, the ratio 
becomes negative, as happens for real meta materials. Again, these need to be analyzed more carefully with the inclusion of backreaction, to come to a 
definitive conclusion. 

Before we end this section, we remind the reader that as we have mentioned, there are various caveats in this treatment, which needs to be analyzed more carefully. 
We leave this for a future study. We emphasize that we make no claims beyond the statement that higher derivative corrections might introduce important changes in the optical response 
of strongly coupled boundary theories. 

\section{\textbf{Discussions and Conclusions}}

In this paper, we have studied generalized holographic superconductors with higher derivative interactions 
in four dimensional AdS-Schwarzschild backgrounds. This generalizes the model considered in \cite{Franco} and \cite{Siopsis}.  We found rich phase structure in the space 
of the coupling strength $\eta$ and the model parameter $\Sigma$. Interestingly, our numerical analysis indicates the presence of two critical $\eta$'s, namely $\eta_{c1}$ and $\eta_{c2}$, 
which form a window inside which the transition from the normal to the superconducting phase is of first order, and away from this window the phase transition is of second order. The 
dependence of $\eta_{c1}$ and $\eta_{c2}$ on $\Sigma$ and $\kappa$ have been established. For fixed $\eta$, we also found a critical $\Sigma_{c}$ around which the nature of phase 
transition changes, but this case is qualitatively different from the one with fixed $\Sigma$. These numerical results were substantiated by studying the free energy
of the boundary theory. The result is shown in fig.(\ref{fullphasestructure}).

It is important to point out the differences of our model with the ones considered in \cite{Franco}. Specifically, the rich phase structure in our model is due to the generalized 
higher derivative coupling term with parameters $\eta$ and $\Sigma$, rather than the a generalized minimal coupling term. Here, we worked with the full backreacted solution, 
and found that backreaction makes the condensation harder to form, and for fixed 
non-zero $\eta$ and $\Sigma$, backreaction effects can also tune the the order of the phase transition. 

We also analyzed the optical conductivity of boundary superconducting system by varying the model parameters $\Sigma$ of the theory as well as the coupling parameter $\eta$. 
Large variations in the ratio of $\omega_g/T_c$ was observed. We further presented analytic results for the critical temperature and showed that these match well with numerical 
analysis, in appropriate regions of the parameter space. 
Finally, we discussed optical properties of the boundary theory with higher derivative corrections, in the probe limit. Our results here are indicative of the fact that such corrections might introduce
important differences in physical quantities like refractive index and magnetic permeability. However, in order to firmly establish these results, one needs to resort to an analysis with
backreaction effects. We leave this for a future study.  

\begin{center}
\bf{Acknowledgements}
\end{center}
It is a pleasure to thank Sayantani Bhattacharyya for useful comments. The work of SM is supported by grant no. 09/092(0792)-2011-EMR-1 from CSIR, India.

\end{document}